\documentclass[a4paper,fleqn]{cas-dc}

\usepackage[backref=false,backend=biber,style=numeric-comp,sorting=none,maxcitenames=1,block=none,giveninits,url=false]{biblatex}
\usepackage{placeins}
\usepackage{subcaption}
\usepackage{graphicx}
\usepackage{hyperref}
\usepackage{float}
\usepackage{xcolor}
\usepackage[export]{adjustbox}

\def\tsc#1{\csdef{#1}{\textsc{\lowercase{#1}}\xspace}}
\tsc{WGM}
\tsc{QE}

\begin{document}
\let\WriteBookmarks\relax
\def\floatpagepagefraction{1}
\def\textpagefraction{.001}

\shorttitle{}

\shortauthors{}

\title [mode = title]{Construction and Commissioning of the Warwick TPC Platform}

\author[1]{Matthew Snape}[orcid=0009-0003-4044-8440]
\ead{Matt.Snape@warwick.ac.uk}

\author[1]{Hong-Kai Tan}[orcid=0009-0000-4873-9638]
\ead{Hong-Kai.Tan@warwick.ac.uk}

\author[1]{Gary Barker}

\author[1]{Alan Burton}

\author[2]{Alexander Deisting}[orcid=0000-0001-5372-9944]
\cormark[1]
\ead{alexander.deisting@cern.ch}

\author[3]{Philip Hamacher-Baumann}

\author[1]{Xianguo Lu}[orcid=0000-0002-3077-1805]
\cormark[1]
\ead{Xianguo.Lu@warwick.ac.uk}

\cortext[1]{Corresponding authors}

\affiliation[1]{
    organization={Department of Physics, University of Warwick},
    city={Coventry},
    postcode={CV4 7AL},
    country={United Kingdom}
}

\affiliation[2]{
    organization={Institut f{\"u}r Physik \&
                  Exzellenzcluster PRISMA$^{+}$,
                  Johannes Gutenberg-Universit{\"a}t Mainz},
    city={Mainz},
    postcode={55128},
    country={Germany}
}

\affiliation[3]{
    organization={AIXTRON},
    country={Germany}
}

\begin{abstract}
Active-target gas time projection chambers (TPCs) could provide the ultimate measurements of neutrino interactions needed for future neutrino oscillation measurements. 
Their development requires optimisation of the gas mixture, pressure, electric fields, and readout configuration. 
 Warwick TPC (WarTPC) is an R\&D platform for investigating in these directions  
 for neutrino physics and beyond. It can operate at pressures up to approximately 10~bar and 
 features a combination of thick gas electron multipliers 
 and  a Timepix3 camera for optical readout.  In this paper, we present the construction and commissioning of the WarTPC platform, describing the design and integration of its subsystems, and report the  
 first measurements of its optical response under different operating conditions, demonstrating a stable and  functioning R\&D platform.

\end{abstract}


\maketitle

\tableofcontents 

\section{Introduction}
\label{sec:1-intro}

Neutrino oscillation measurements in the GeV regime require precise knowledge of neutrino interactions to reconstruct the neutrino flux and energy. 
This is  
complicated by the underlying interaction dynamics and intranuclear medium effects. Gas time projection chambers (TPCs) can be used as active-target   detectors for neutrinos. They offer low detection thresholds and a 4$\pi$ angular acceptance that are crucial for 
understanding these effects. 
Their immediate drawback---limited target mass---can be compensated by using a large detector volume with pressured gas in an intense neutrino beam. 
At a pressure of 10~bar, a gas medium still performs better in tracking than solid or liquid~\cite{TKI-HPgTPC}. 
 TPC can operate with a range of gas mixtures; hydrogen-rich ones could enable direct measurements of neutrino  interactions on hydrogen and hence  neutrino energy reconstruction free of nuclear effects~\cite{TKIpaper,TKI-HPgTPC}.

A sufficient target mass in an active-target TPC for neutrinos requires a large detector volume and a capable readout system. Conventional charge readout can be costly as the number of channels increases with the readout area. 
Optical readout offers a potentially scalable alternative, with a small sensor imaging a much larger area while providing fine spatial granularity and good timing resolution~\cite{ARIADNE-demo}. 
 In noble gases such as argon and xenon, ionisation electrons can produce secondary scintillation, or electroluminescence~\cite{Neutral_bremsstrahlung_and_excimer_electroluminescence_in-noble_gases_and_its_relevance_to_two_phase_dark_matter_detectors}. The corresponding optical tracking signal can be imaged by cameras.

The electron-transport properties and light yield that determine TPC performance rely on the gas composition, pressure, and drift and amplification fields.  
It may not be possible to achieve  suitable electron-transport properties and sufficient light yield simultaneously in hydrogen-rich mixtures at high pressure within practical limitations such as high voltage and drift length. 
In particular,  
methane can suppress noble-gas scintillation~\cite{Henriques2019}, while increasing pressure can reduce the light yield~\cite{Neutral_bremsstrahlung_and_excimer_electroluminescence_in-noble_gases_and_its_relevance_to_two_phase_dark_matter_detectors}. 
It is therefore important to verify such expected limitations \textit{in situ} and identify suitable combinations of gas mixture, pressure, electric fields, and readout configuration. 
Hence, an R\&D platform is required. 
Such a platform may also reveal further opportunities for active-target, or general tracking, TPCs through a systematic exploration of operating conditions, including, for example, at low pressures or with charge readout.

Warwick TPC  (WarTPC) is such a platform for investigating novel TPC technologies for neutrino physics and beyond. Construction began in August 2022, and 
commissioning has been concluded with the observation of alpha-particle tracks from a 
$^{241}$Am  source and cosmic-ray tracks. 
This paper presents an overview of the construction and commissioning of WarTPC,  describing the design and integration of its subsystems and the first optical signal observations under different operating conditions. More details can be found in Ref.~\cite{SnapeThesis}. 
This paper is organised as follows. Section 2 describes WarTPC and its subsystems. Section 3 describes the commissioning measurements, followed by the conclusions in Section 4.

\section{Construction}
\label{sec:2-wartpc}

\begin{figure*}
    \centering
    {\includegraphics[width=0.8\textwidth]{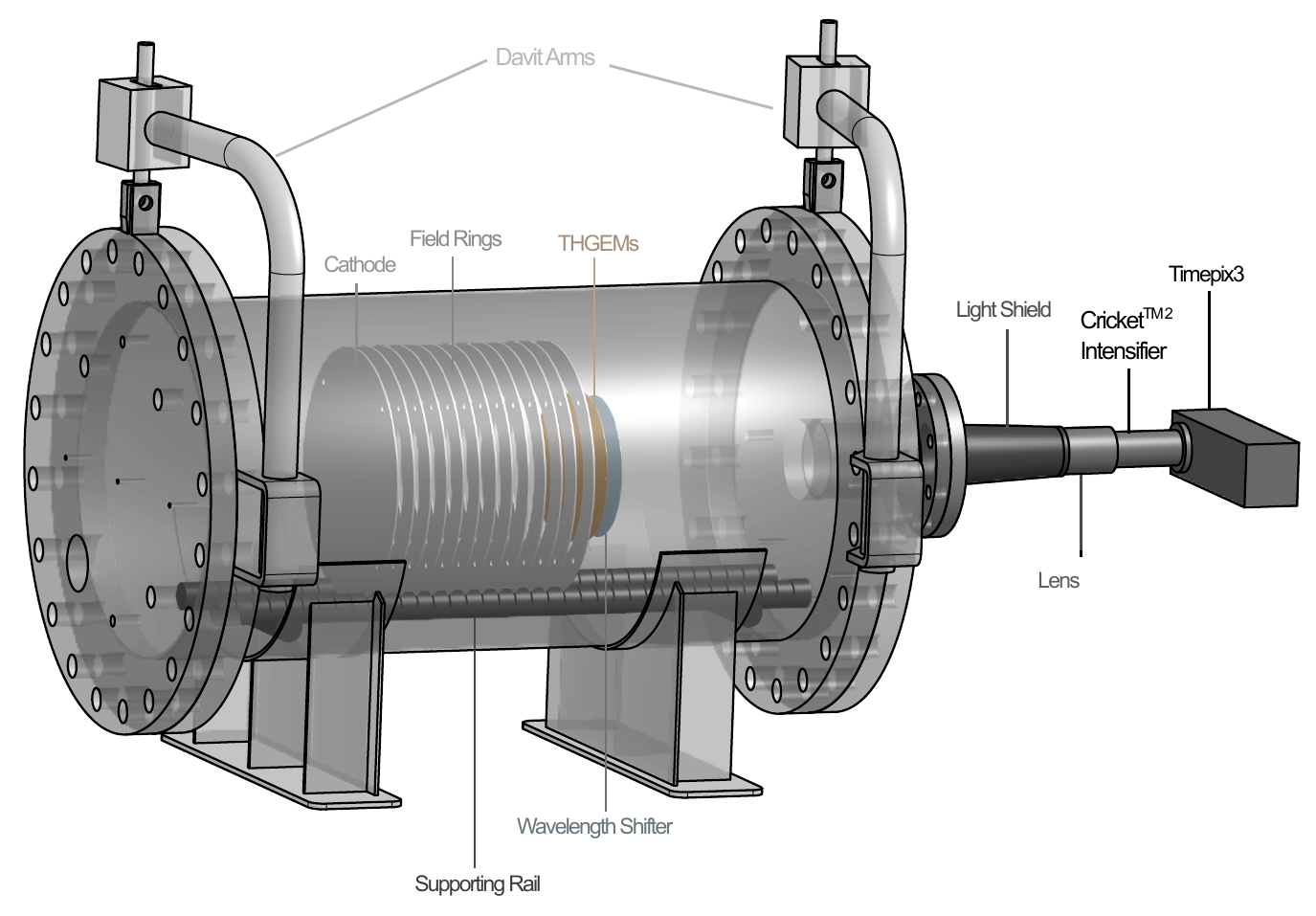}}
    \caption{CAD schematic of the WarTPC platform, showing the cathode, field rings and their supporting rail, double-THGEM stack, and wavelength shifter inside the pressure vessel (not labelled), together with the external camera system consisting of the light shield, lens,  intensifier, and Timepix3 camera. The camera views the wavelength shifter through the vessel viewport. The davit arms supporting the vessel doors are labelled. The gas supply and the supporting systems are not shown.}
    \label{fig:setupcad}
\end{figure*}

\begin{table*} 
    \centering
    \caption{\centering The WarTPC supporting systems}
    \label{tab:components}
    \begin{tabular}{lll}
        \toprule
        \textbf{System} & \textbf{Component} & \textbf{Function} \\
        \midrule
        
        \multirow{5}{*}{Slow control} 
          & Wika A-10 Pressure Sensor & Pressure monitoring \\
          & 2 $\times$ Carel 10 kOhm NTC Temperature Probes & Temperature monitoring \\
          & Keithley DAQ6510 & Slow control logging and display \\
          & Hiden QGA residual gas analyser & Gas composition analysis \\
          & Inficon PSG550 \& PGD500 vacuum gauge and display & Sampling system pressure monitoring \\
        \midrule
        
        \multirow{2}{*}{DAQ} 
          & PicoScope 5244D & Charge readout digitisation \& streaming \\
          & Timepix3 SPIDR Readout Board & Optical readout streaming \\
        \midrule
        
        \multirow{2}{*}{HV} 
          & Caen N570 & Field cage power supply \\
          & Caen N1471H & THGEM power supply \\
        \midrule
        \multirow{4}{*}{Electronics} 
          & 2 $\times$ Cremat CR-110 R2.2 CSPs & Charge signal integration/amplification \\
          & 2 $\times$ CR-150-R6 CSP Evaluation Boards & Charge signal decoupling\\
          & Photonis Cricket 2 Intensifier & Optical signal amplification\\
          & Timepix3 Chip & Optical readout\\
        \bottomrule
    \end{tabular}
\end{table*}

\begin{figure*}
  \centering
    \subfloat[\label{fig:fullcathode}]{%
    \adjincludegraphics[
      trim={{.1\width} {.25\height} {.3\width} {.3\height}},
      clip,
      width=0.85\columnwidth
    ]{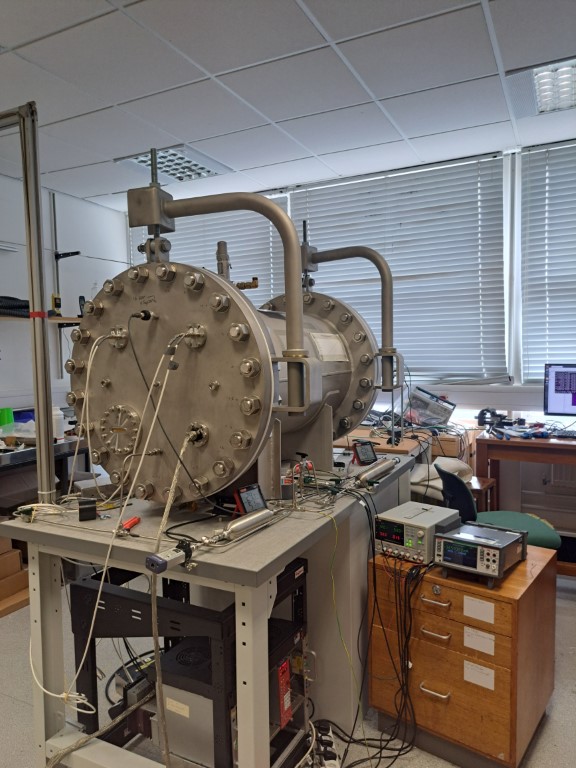}}\hfill
  \subfloat[\label{fig:fullcamera}]{%
    \adjincludegraphics[
      trim={0 {.2\height} {.07\width} {.3\height}},
      clip,
      width=1.19\columnwidth
    ]{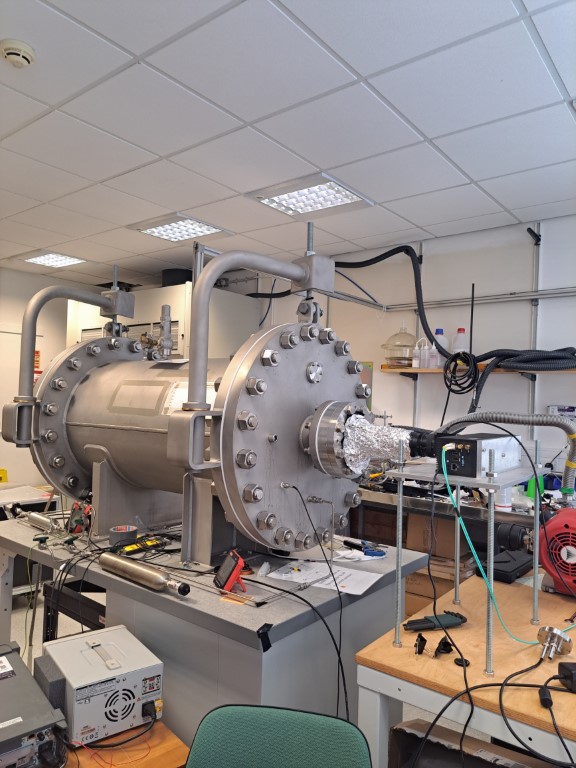}}

  \caption{The fully integrated WarTPC platform viewed from (a) the cathode side and (b) the camera side in August 2026.}
  \label{fig:wartpc-full-setup}
\end{figure*}

WarTPC is an R\&D platform for pressure and gas-mixture studies in an active-target gas TPC, with the capability to test different readout technologies.  Figure~\ref{fig:setupcad} shows the detector schematic, including the 195~L pressure vessel, a TPC with an approximately 30~cm drift length, an amplification region  with an active area of $\diameter$175~mm, and a Timepix3-based camera system. The amplification region consists of a double-thick gas electron multiplier (THGEM) stack and a wavelength shifter capable of producing both charge and optical signals. 
The supporting systems, including slow control, a  data acquisition (DAQ) system, high-voltage (HV) power supplies, and electronics, are summarised in Table~\ref{tab:components}. 
The fully integrated platform is shown  in Fig.~\ref{fig:wartpc-full-setup}. The following subsections describe these components and their integration.
 
\subsection{Pressure Vessel}

\begin{figure}
  \centering
  \subfloat[Trial fitting on the support table.]{%
    \includegraphics[width=0.49\columnwidth]{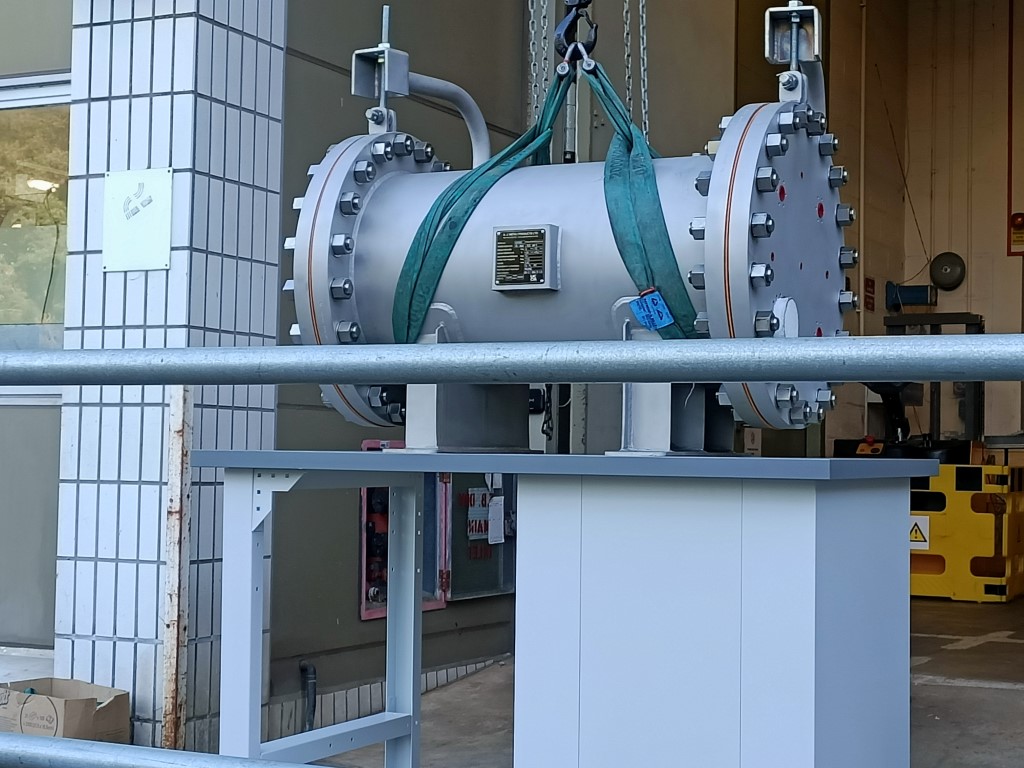}}
  \hfill
  \subfloat[Positioning the vessel in the laboratory.]{%
    \includegraphics[width=0.49\columnwidth]{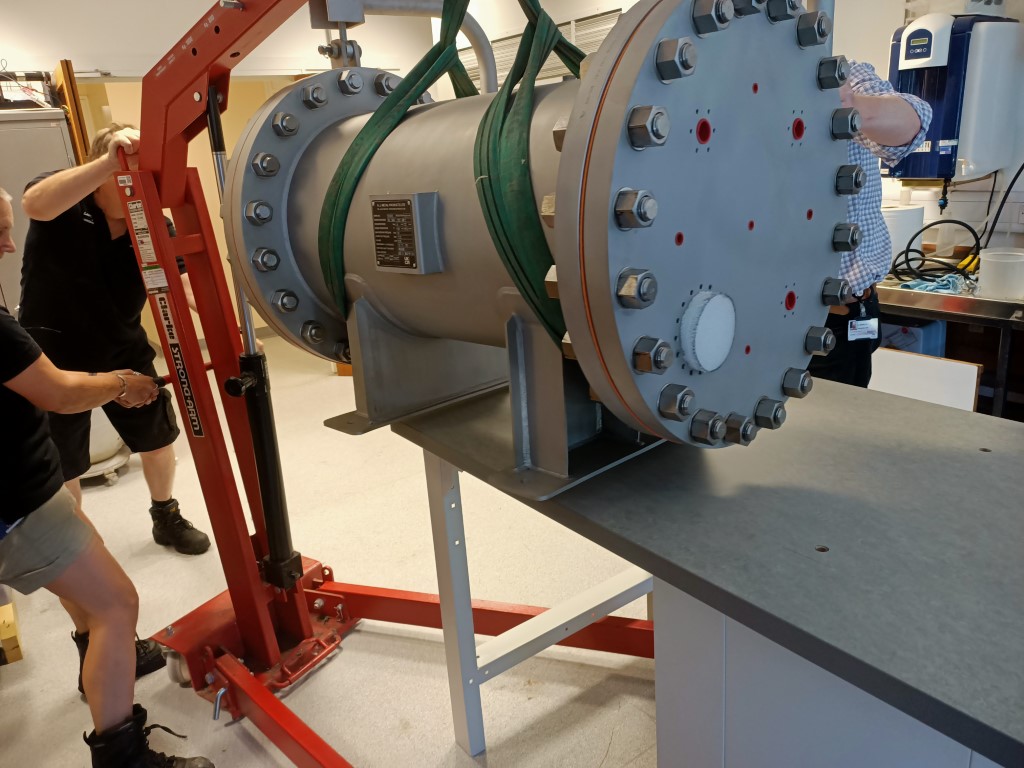}}

  \par\medskip
  \subfloat[Side view of the vessel secured to the support table.]{%
    \adjincludegraphics[trim={{0.08503958\width} {0.16063031\height} {0.14173263\width} {0.00000000\height}},clip,width=0.9\columnwidth]{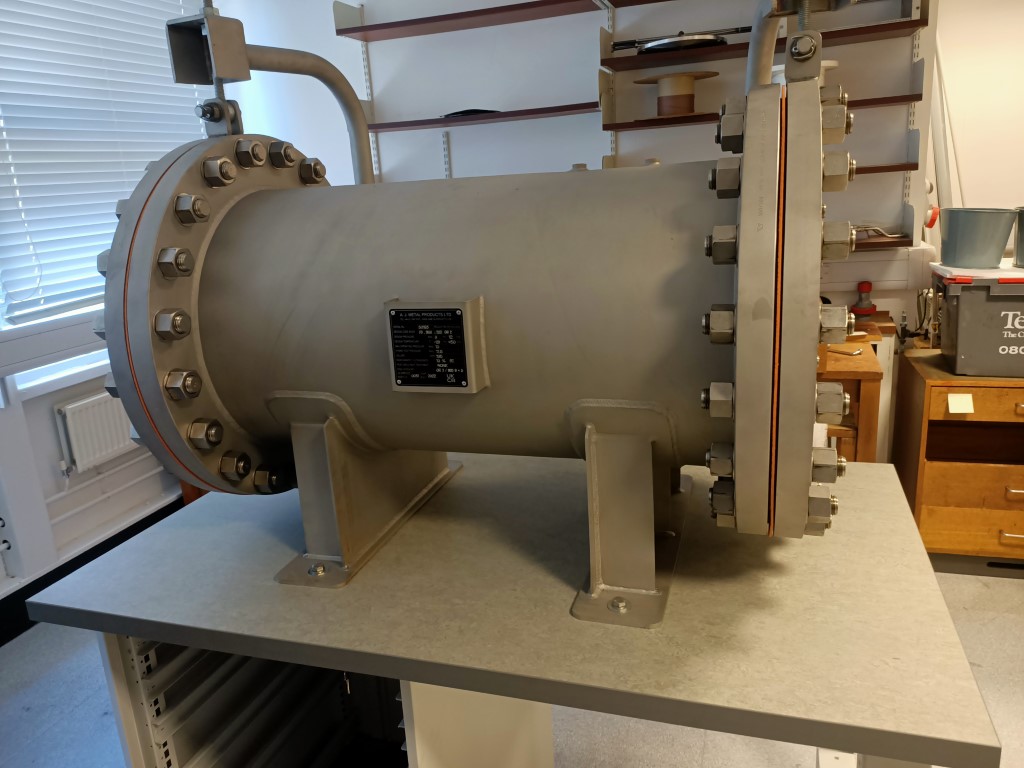}%
    \label{fig:vessel-length-ways}}
  \caption{Installation of the pressure vessel in August 2022.}
  \label{fig:initial-assembly-of-vessel-and-table}
\end{figure}

The pressure vessel (Figs.~\ref{fig:setupcad} and~\ref{fig:vessel-length-ways}) is cylindrical, with an outer diameter of 508~mm, a wall thickness of 5~mm, and a distance of 1~m between the raised faces of its end flanges. Each end has a DN500 raised-face slip-on (RFSO) flange with 20 clearance holes for M30 bolts on a pitch circle diameter (PCD) of 650~mm. The vessel has a capacity of 195~L and weighs 580~kg excluding the doors, which weigh approximately 150~kg each. The vessel body, davit arms, and doors are made from 316L stainless steel.

The vessel has a design pressure of 12~barg and a normal operating pressure of 10~barg. It passed a hydrostatic test at 17.5~barg for 30~minutes on 28 March 2022. Two tapped 1/4-inch BSPT ports at the top of the vessel host a pressure relief valve and a pressure gauge. The relief valve is set to open at 10~barg to protect against overpressure.

The vessel was trial-fitted on the support table using a crane on the ground floor, where the mounting holes were marked out and drilled. The table, which can support a load of 2~tonnes, was then moved into the laboratory. The vessel was lifted onto it using a mobile hydraulic crane and secured with eight M10 bolts, four on each saddle support. The installation is shown in Fig.~\ref{fig:initial-assembly-of-vessel-and-table}. 

\begin{figure}
  \centering
  \subfloat[Before grinding.]{\includegraphics[width=0.44\columnwidth]{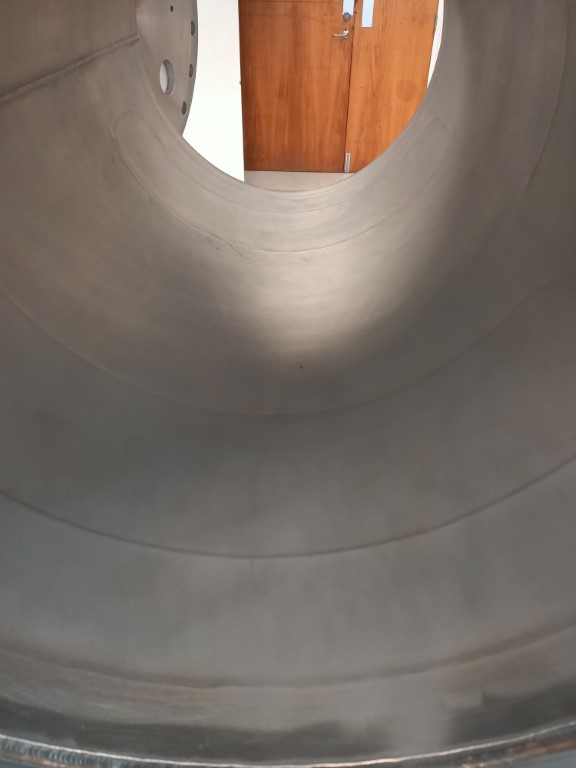}\label{fig:vessel-before-smoothing}}
  \hfill
  \subfloat[After grinding.]{\adjincludegraphics[trim={{0.00000000\width} {0.17716578\height} {0.00000000\width} {0.00000000\height}},clip,width=0.53\columnwidth]{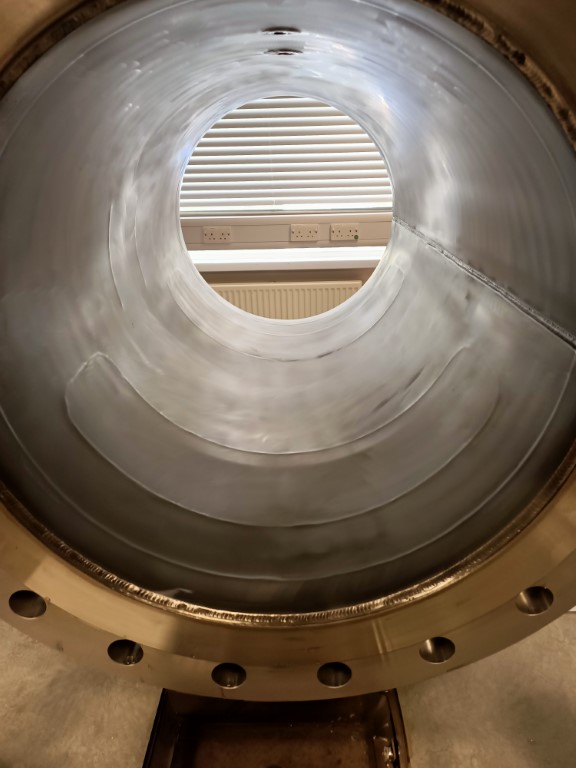}\label{fig:vessel-after-smoothing}}
  \caption{Inner surface of the pressure vessel before and after grinding to remove rough areas.}
  \label{fig:vessel-grinding}
\end{figure}

During cleaning with polyester/cellulose wipes soaked in isopropyl alcohol (IPA), the inner surface was found to be rough enough to snag the wipes. The surface was therefore smoothed using an angle grinder, as shown in Fig.~\ref{fig:vessel-grinding}.

\subsubsection{Vessel Doors}\label{subsubsec:vesseldoors}

\begin{figure}
  \centering
  \subfloat[Cathode-side door.]{\adjincludegraphics[trim={{0.00000000\width} {0.09448842\height} {0.17716578\width} {0.11338610\height}},clip,width=0.47\columnwidth]{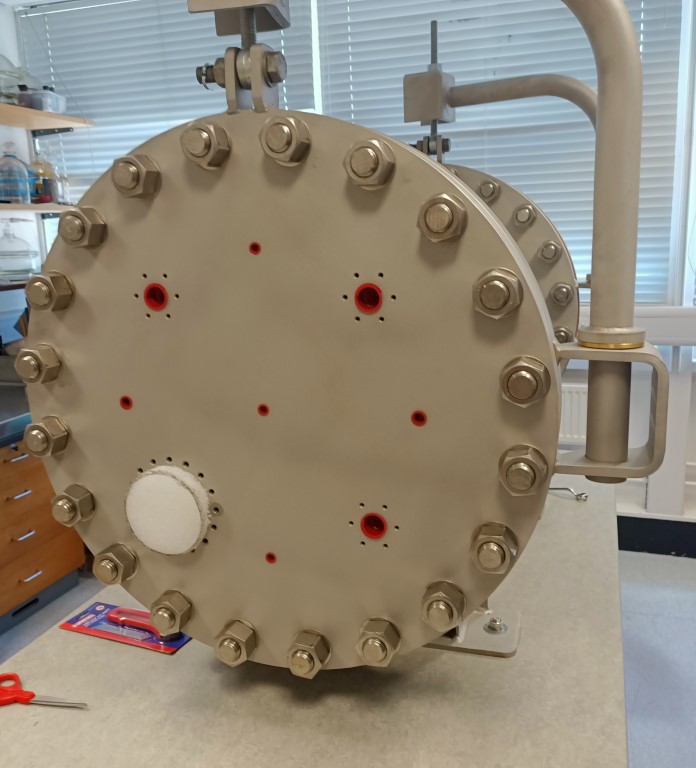}\label{fig:vessel-anode-side}}
  \hfill
  \subfloat[Camera-side door.]{\adjincludegraphics[trim={{0.05210758\width} {0.09448842\height} {0.03126455\width} {0.07559073\height}},clip,width=0.50\columnwidth]{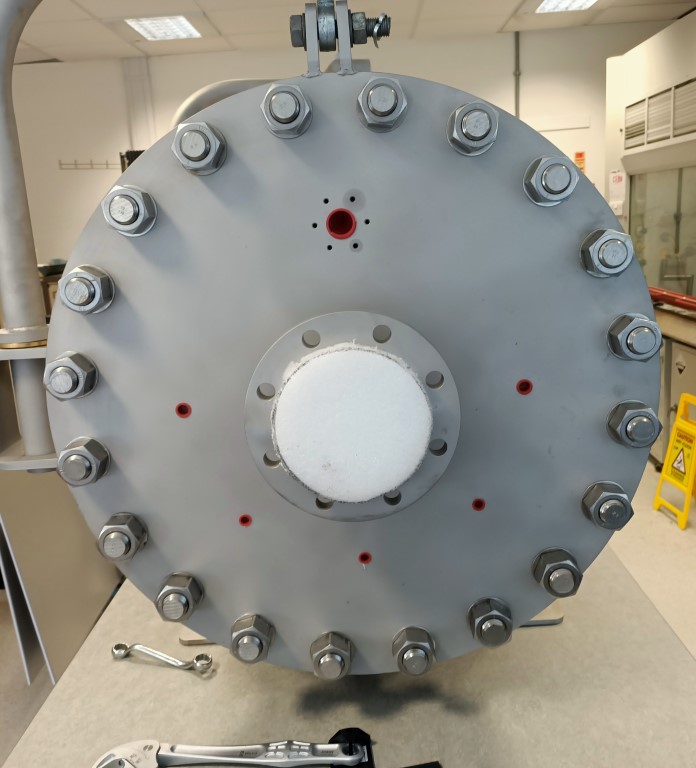}\label{fig:vessel-cam-side}}
  \par\medskip
  \subfloat[Bolt-tightening sequence.]{\includegraphics[width=0.49\columnwidth]{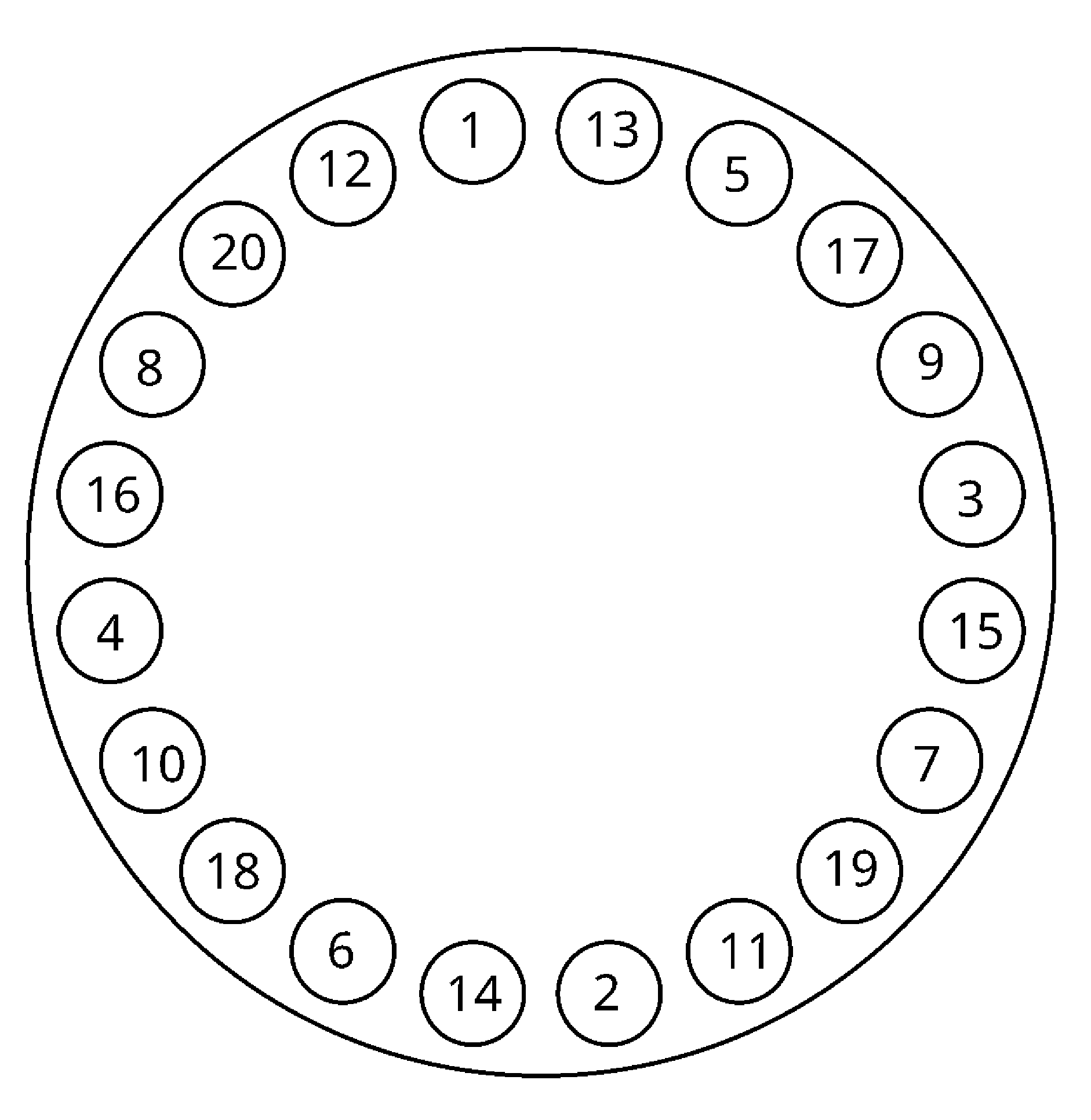}\label{fig:bolt-tightening-order}}
  \caption{Vessel doors and bolt-tightening sequence. The Y, Z, and W holes have red protective linings, while the X hole and N1 nozzle are protected by foam inserts.}
  \label{fig:vessel-holes}
\end{figure}

The vessel opens at both ends, with each door supported by a davit arm. The cathode-side door (Figs.~\ref{fig:fullcathode} and~\ref{fig:vessel-anode-side}) primarily hosts feedthroughs for detector electronics, while the camera-side door (Figs.~\ref{fig:fullcamera} and~\ref{fig:vessel-cam-side}) has specific design for the optical readout. Each door is 715~mm in diameter and 44~mm thick, with M30 clearance holes matching those on the DN500 vessel flanges.

A 3~mm-thick PTFE gasket seals each door to the vessel, with clearance holes for the M30 stud bolts. The original compressed non-asbestos fibre (CNAF) gaskets, visible in Fig.~\ref{fig:vessel-length-ways}, were replaced with PTFE gaskets, which deform under compression to conform to the sealing surfaces.

The doors are secured using M30 stud bolts and nuts. A Norbar 4-AR adjustable torque wrench is used to tighten the nuts in three stages: 200, 345, and 567~N\,m. At each stage, opposite pairs are tightened in the sequence shown in Fig.~\ref{fig:bolt-tightening-order} to distribute the load evenly.

\subsubsection{Feedthroughs} 

Both doors have three types of holes, designated W, Y, and Z. The Y holes have 1/4-inch NPT threads for gas connections and vessel monitoring components. The Z holes are tapped M16 and provide additional mounting points, although they are currently unused. Each W hole has a diameter of 30~mm and is surrounded by six equally spaced tapped M6 holes on a pitch circle diameter (PCD) of 58.7~mm. These ports are primarily used for electrical feedthroughs.

The cathode-side door (Figs.~\ref{fig:fullcathode} and~\ref{fig:vessel-anode-side}) has a central Y hole, with two further Y holes, two Z holes, three W holes, and one X hole arranged on a 400~mm PCD. The X hole has a diameter of 80~mm and is surrounded by sixteen equally spaced tapped M8 holes on a 130.25~mm PCD.

\begin{figure}
  \centering
  \subfloat[Feedthrough fitted with the PTFE sleeve.]{%
    \adjincludegraphics[trim={{0.19842568\width} {0.12755936\height} {0.24566989\width} {0.22677220\height}},clip,
      width=0.45\columnwidth]{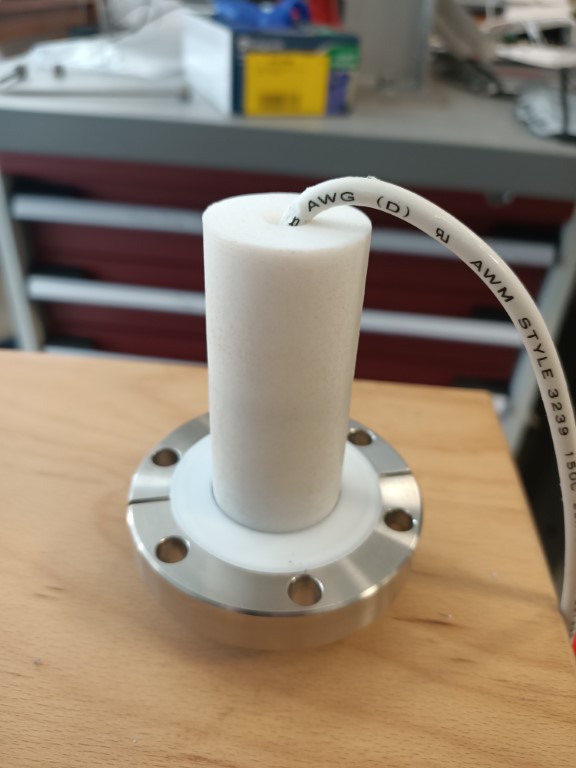}%
    \label{fig:50kV-with-sleeve}}
  \hfill
  \subfloat[View from inside the vessel after installation on the cathode-side door.]{%
    \includegraphics[width=0.52\columnwidth]{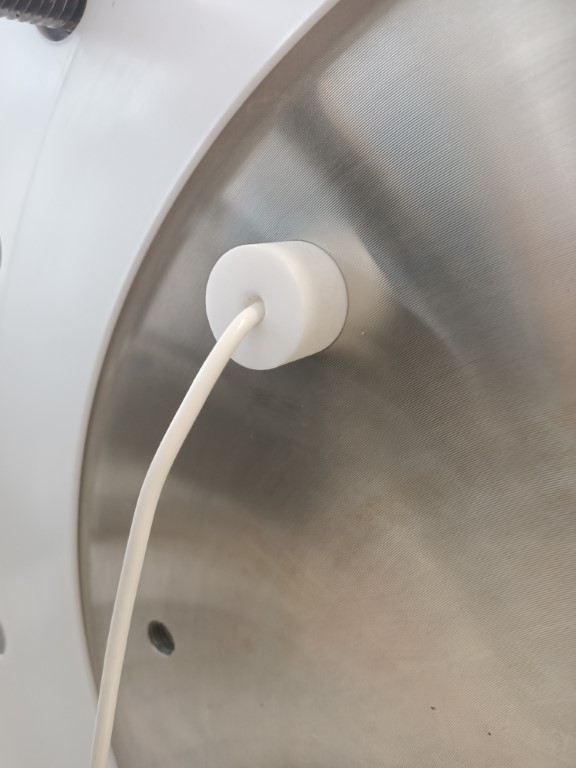}%
    \label{fig:50kV-in-vessel}}

  \caption{Field cage high-voltage feedthrough and its PTFE sleeve before and after installation.}
  \label{fig:50kv-feedthrough-and-ptfe-sleeve}
\end{figure}

The three W ports on the cathode-side door accommodate the high-voltage feedthroughs: one for the field cage and two for the THGEMs. The field cage feedthrough is a custom-made single-pin unit rated to 50~kV DC and 10~A. A PTFE sleeve surrounds its internal connector to reduce the risk of sparking to the vessel door (Fig.~\ref{fig:50kv-feedthrough-and-ptfe-sleeve}). The two THGEM feedthroughs are two-pin SHV units rated to 5~kV DC and fitted with similar PTFE insulating sleeves.

The camera-side door (Figs.~\ref{fig:fullcamera} and~\ref{fig:vessel-cam-side}) has one W, two Z, and three Y holes arranged on a 400~mm PCD. A central nozzle, designated N1, terminates in an RFSO flange and hosts the camera viewport.

\begin{figure}
  \centering
  \subfloat[Viewport assembly.]{%
    \adjincludegraphics[trim={{0.09448842\width} {0.21968557\height} {0.12283494\width} {0.13464600\height}},clip,
      width=0.8\columnwidth]{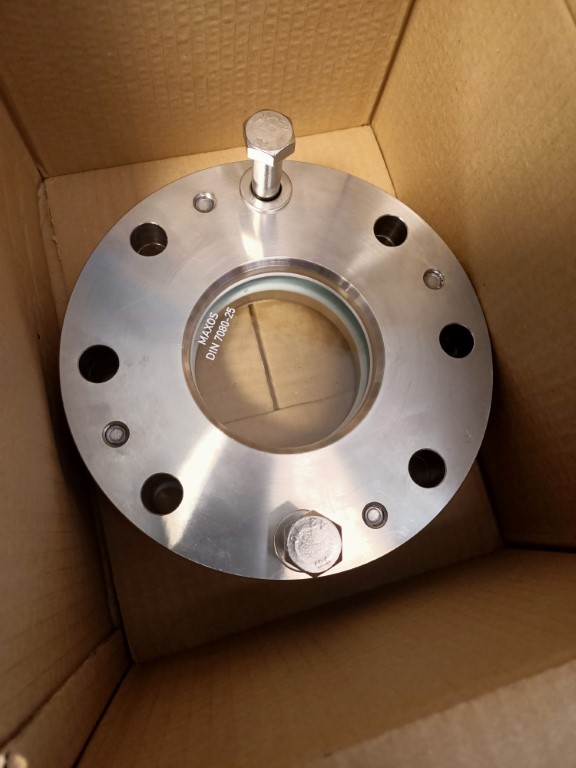}%
    \label{fig:camera-flange}}

  \par\medskip
  \subfloat[Before installation, with the PTFE sealing gasket visible.]{%
    \adjincludegraphics[trim={{0.14173263\width} {0.24803210\height} {0.37795367\width} {0.17716578\height}},clip,
      width=0.35\columnwidth]{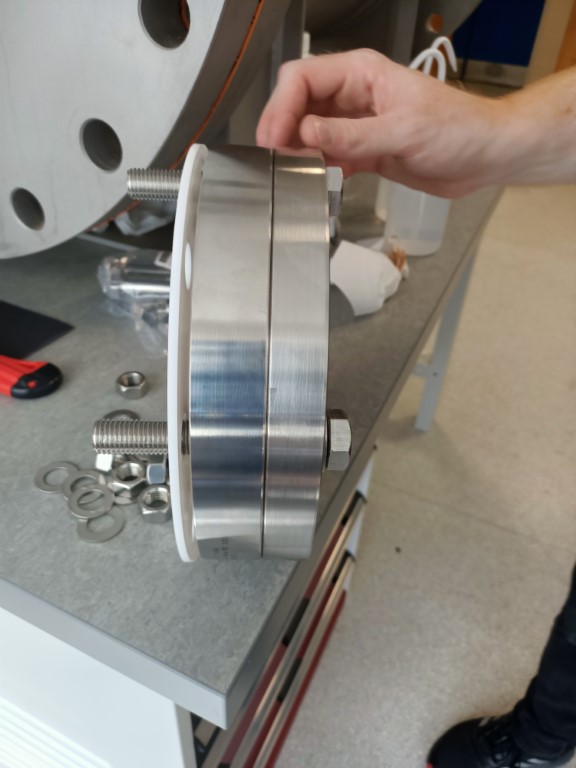}%
    \label{fig:camera-flange-with-gasket}}
  \hfill
  \subfloat[Securing the assembly to N1 with M16 bolts.]{%
    \adjincludegraphics[trim={{0.00000000\width} {0.31889841\height} {0.09448842\width} {0.07086631\height}},clip,
      width=0.62\columnwidth]{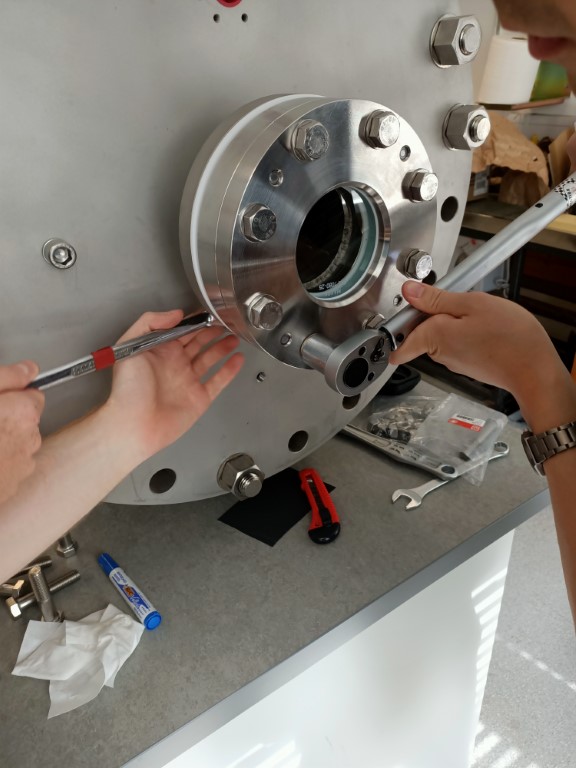}%
    \label{fig:camera-flange-on-vessel}}

  \caption{Camera viewport assembly and its installation on the vessel.}
\end{figure}

The viewport assembly (Fig.~\ref{fig:camera-flange}) consists of a 55~mm-thick 316L stainless steel flange and a toughened borosilicate glass disc with a viewing diameter of 100~mm and a maximum pressure rating of 25~barg. An inner PTFE envelope gasket and an outer compressed-fibre gasket seal the glass within the flange. Eight M16 bolts secure the assembly to N1, with a separate PTFE gasket providing the gas-tight seal between the flanges. Figures~\ref{fig:camera-flange-with-gasket} and~\ref{fig:camera-flange-on-vessel} show the assembly before and during installation.

Unused ports are sealed. The W port on the camera-side door is closed with a blind flange. The central Y port on the cathode-side door is fitted with a Swagelok adapter from a 1/4-inch male NPT thread to a 6~mm tube connection, which is sealed with a Swagelok 6~mm plug. The Z holes are sealed with M16 bolts with PTFE tape wrapped around their threads. As these holes are tapped through the full door thickness, they can be sealed from either side.

The X port is fitted with a 6-inch (152.4~mm) CF flange incorporating a 50-pin connector, reserved for future connections to miscellaneous electronics or sensors.

\subsection{Slow Control}\label{subsubsec:vesselmonitoring}

The slow control system monitors the vessel pressure, temperature, and gas supply.

\subsubsection{Vessel Monitoring}

The vessel pressure is measured by a WIKA A10 pressure sensor installed in one of the Y holes on the cathode-side door and  powered by a 12~V DC supply. Its output current is converted to pressure using the calibration curve supplied with the sensor.

Two Carel 10~k$\Omega$ negative temperature coefficient (NTC) probes measure the temperature inside the vessel. Each probe is secured using a Swagelok nut and ferrule to a 10~mm-to-1/4-inch NPT fitting installed in a Y hole. The probe resistance is converted to temperature using the manufacturer's calibration data, which cover $-50\,^{\circ}\mathrm{C}$ to $110\,^{\circ}\mathrm{C}$. Within the typical laboratory temperature range of $18$--$28\,^{\circ}\mathrm{C}$, the response is approximately linear.

These sensor currents and resistances are recorded by a Keithley DAQ6510 multichannel digital multimeter.

\subsubsection{Gas System}

\begin{figure}
    \centering
    \includegraphics[width=\columnwidth]{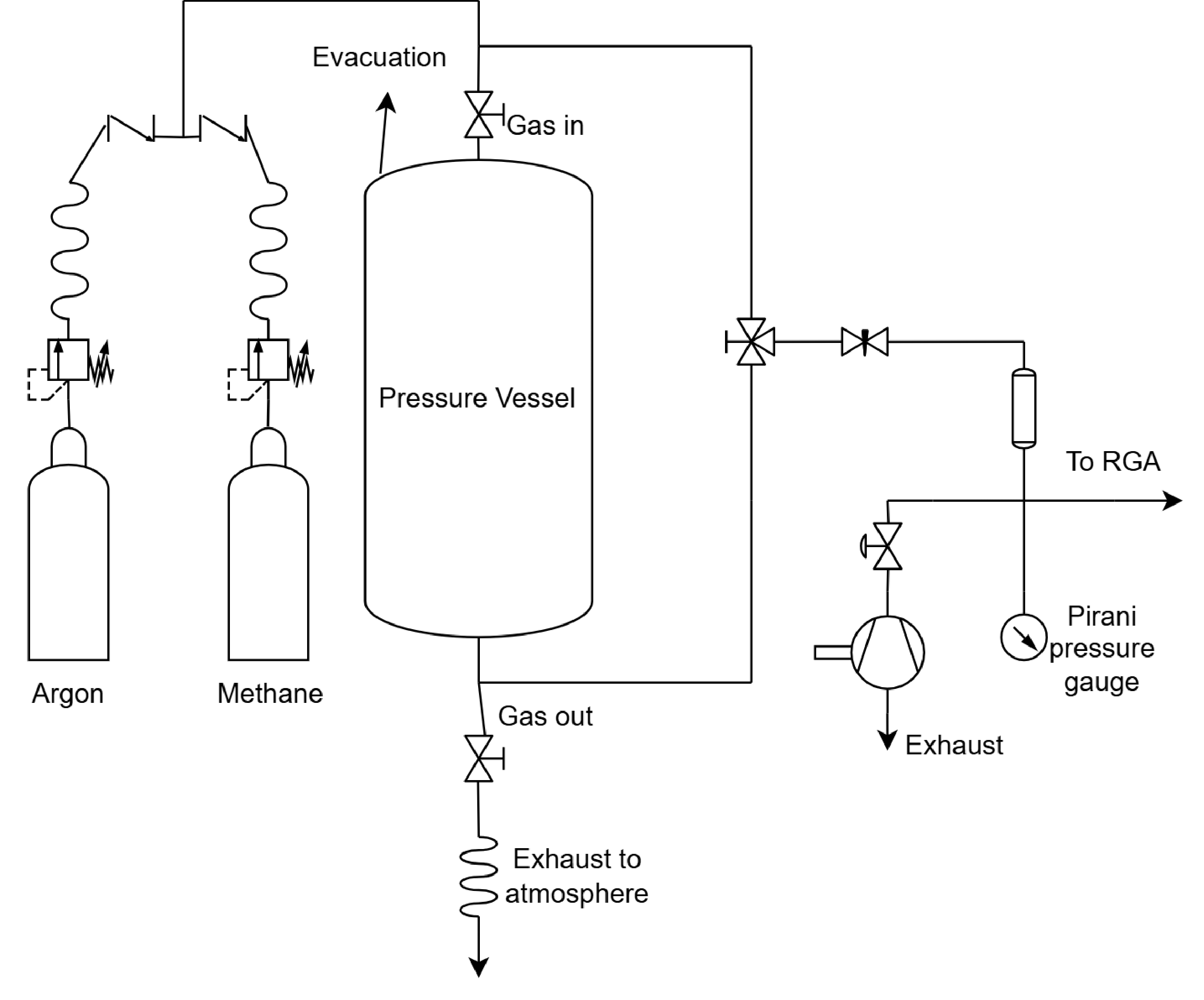}  
   \caption{Schematic of the gas system.}    \label{fig:wartpc_gas_system} 
  \end{figure}

\begin{figure}
  \subfloat[ Gas inlet to the sampling system (A) and valve to the vessel (B). Gas enters at the centre and branches towards A and B.
  \label{fig:gas-system-inlet}]{%
  \adjincludegraphics[trim={{0.33070946\width} {0.35433157\height} {0.00000000\width} {0.35433157\height}},clip,width=0.95\columnwidth]{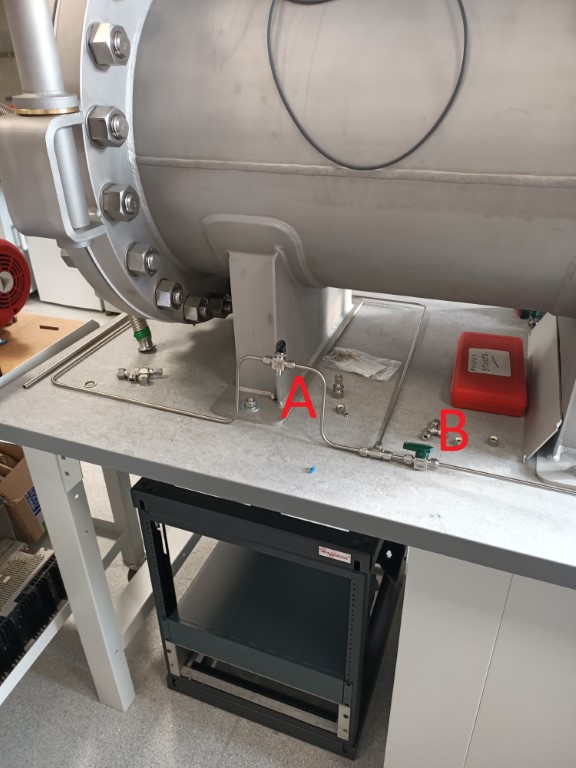}}
  \par\medskip
  \subfloat[
  Gas exhaust valve to the fume cupboard exhaust (A) and from the sampling system (B).\label{fig:gas-system-exhaust}]{%
  \adjincludegraphics[trim={{0.06614189\width} {0.28346525\height} {0.04724421\width} {0.35433157\height}},clip,width=0.95\columnwidth]{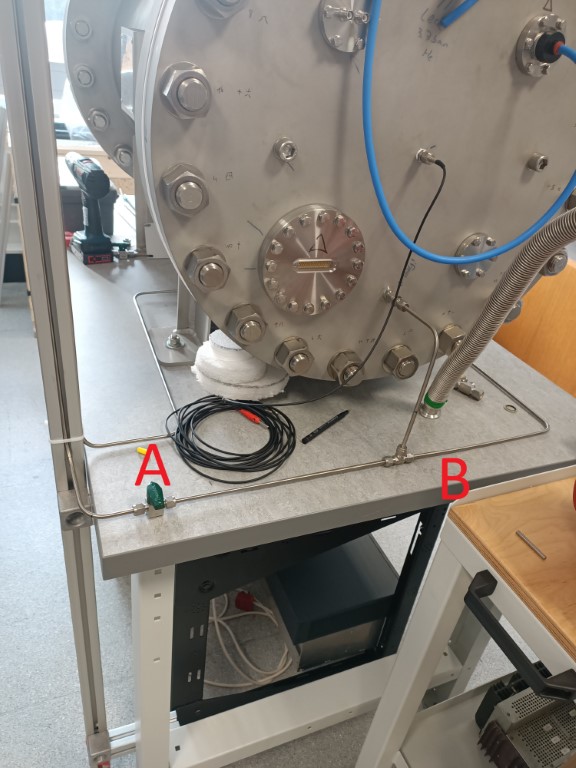}}
\par\medskip
\subfloat[Sampling system with the three-way valve (A), low-flow metering valve (B), sampling bottle (C), RGA inlet (D), vacuum pump flange (E), and low-pressure gauge (F).
  \label{fig:sampling-system}]{
  \adjincludegraphics[trim={{0.00000000\width} {0.09263571\height} {0.00000000\width} {0.09263571\height}},clip,width=0.95\columnwidth]{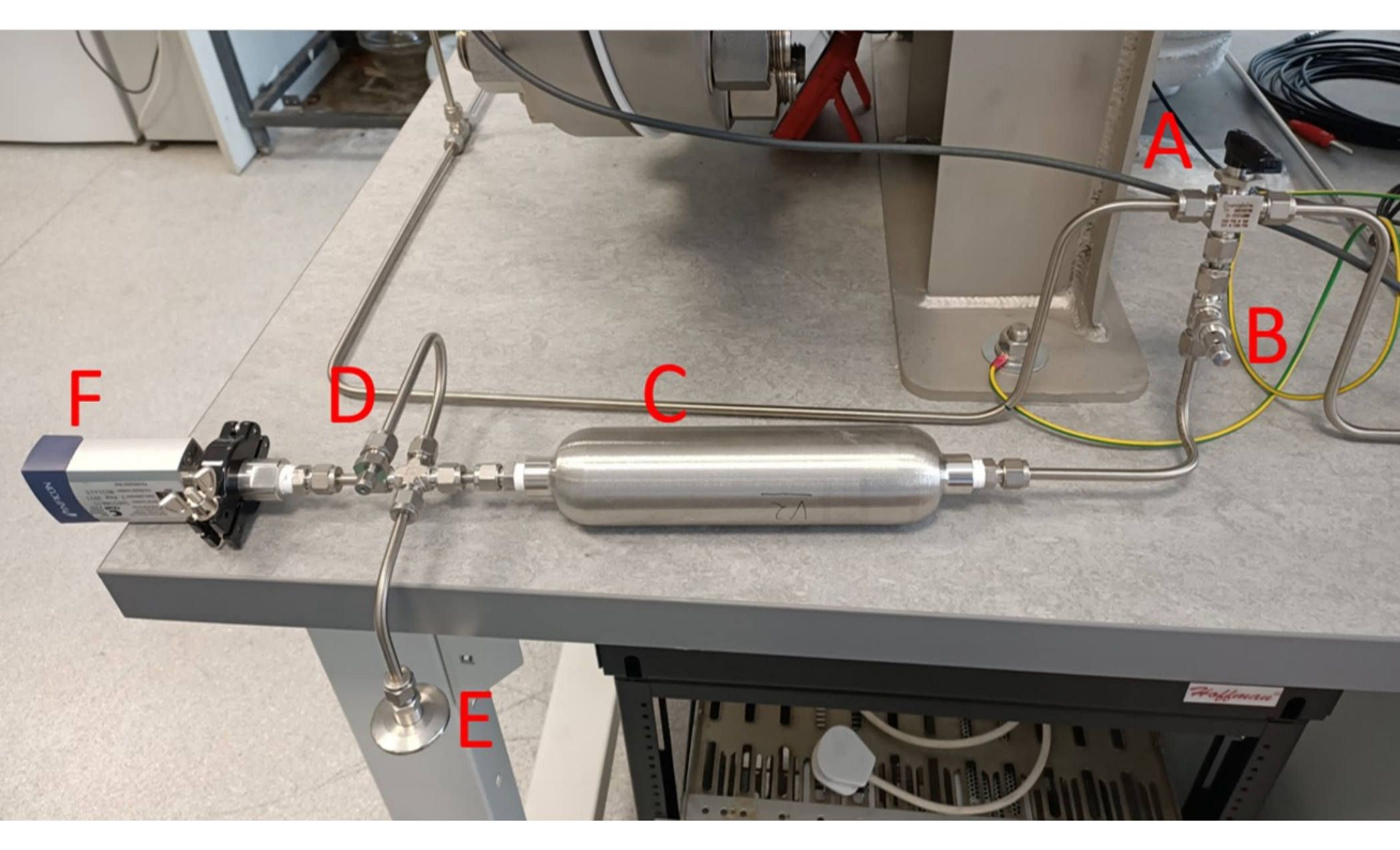}
  }
    \caption{Gas inlet, exhaust, and sampling system.}\label{fig:wholegas}
\end{figure}

The gas system (Fig.~\ref{fig:wartpc_gas_system}) allows operation with argon alone or mixed with a second gas. This work used  argon (grade N6.0), argon--methane mixtures, and argon--CF$_4$ mixtures. The gases mix by diffusion within the vessel and the component fractions are calculated by partial pressure. The gas lines use 6~mm tubing, with the inlet and exhaust arrangements shown in Figs.~\ref{fig:gas-system-inlet} and~\ref{fig:gas-system-exhaust}, respectively.

A separate evacuation line connects the vessel to a Pfeiffer HiScroll 6 pump through a Y hole on the camera-side door. Although the pump has a nominal pumping rate of 6.1~m$^3$/h, the evacuation rate is limited by the 6~mm tubing. The vessel has been tested for a pressure of approximately $10^{-3}$~bar.

The sampling system (Fig.~\ref{fig:sampling-system}) uses a Hiden Analytical QGA residual gas analyser (RGA) to measure the gas composition. A scroll pump and turbomolecular pump maintain the RGA at the required operating pressure of $10^{-7}$--$10^{-6}$~mbar. A three-way valve selects gas from either the vessel or the supply cylinder. Measurements of grade N6.0 argon from the cylinder provide a calibration for the vessel measurement. 

A low-flow metering valve and sampling bottle reduce the gas pressure before it reaches the RGA inlet. Gas enters the RGA through a heated quartz capillary, which evaporates the residual water. An Inficon PSG550 pressure gauge, connected to an Inficon PGD500 display, monitors the sampling-line pressure, while a roughing scroll pump evacuates the line.

\subsection{TPC}

The TPC consists of a field cage, which provides an electric field to drift the ionisation electrons, and an amplification stage, where charge avalanches and secondary scintillation are produced.

\subsubsection{Field Cage}\label{subsubsec:fieldcage}

\begin{figure}
  \centering
  \subfloat[Assembled field cage, with the THGEMs and WLS on the left and the source holder near the centre.
  \label{fig:fc-full-assembly}]{
  \adjincludegraphics[trim={{0.12755936\width} {0.04724421\height} {0.12047273\width} {0.18897684\height}},clip,width=0.9\columnwidth]{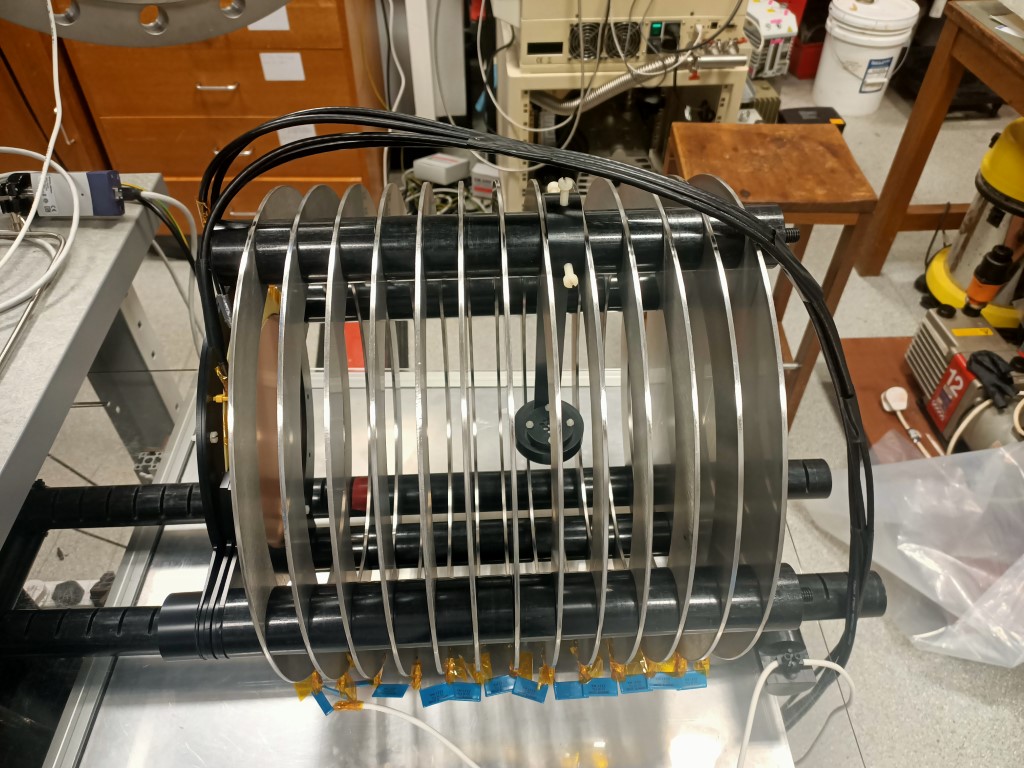}
  }
  \par\medskip
  \subfloat[Measurement of a Delrin spacer.\label{fig:delrin-spacer}]{
  \adjincludegraphics[trim={{0.00000000\width} {0.03543316\height} {0.00000000\width} {0.24803210\height}},clip,width=0.8\columnwidth]{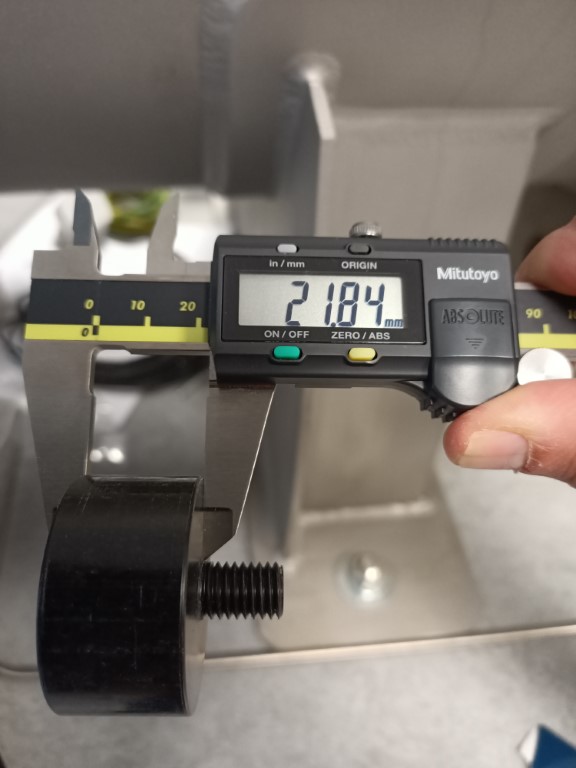}
  }
  \caption{Field cage assembly and a Delrin spacer.}\label{fig:tpcass}
\end{figure}

The TPC field cage (Fig.~\ref{fig:fc-full-assembly}) consists of a cathode and 12 field-shaping rings, with a total drift length of 308~mm. Each ring has an inner diameter of 280~mm and a thickness of 3~mm. The cathode is a circular steel sheet, 360~mm in diameter and 3~mm thick.

Delrin spacers between adjacent rings provide structural support (Fig.~\ref{fig:delrin-spacer}). This modular arrangement allows the field cage to be extended or shortened. The assembled field cage rests on two supporting rails made of Delrin rods with slots at a pitch of 21.8~mm to hold the rings. Two additional Delrin rods underneath raise the assembly in the vessel to align with the viewport.

\begin{figure}
    \centering
    \includegraphics[width=0.95\columnwidth]{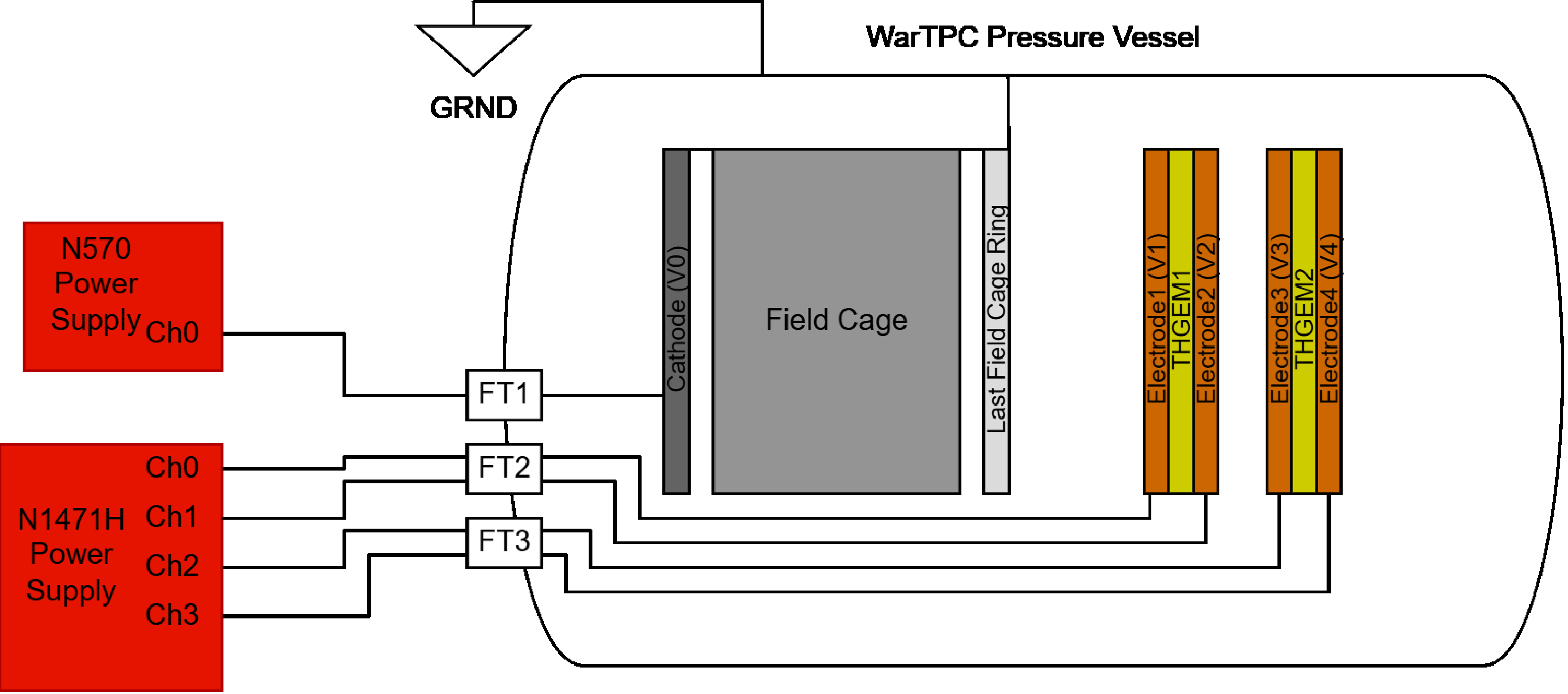}
    \caption{High-voltage power supply connections. FT1 is the field cage feedthrough; FT2 and FT3 are the two-pin SHV feedthroughs.}
    \label{fig:HV_PS-diagram}
\end{figure}

A CAEN N570 15~kV power supply is used to bias the cathode (Fig.~\ref{fig:HV_PS-diagram}). A resistor chain distributes the potential across the rings, with twelve 10~M$\Omega$ resistors connecting the cathode and successive rings. A 50~M$\Omega$ resistor is added between the final ring and ground. With a cathode voltage of $-15$~kV, this divider holds the final ring at approximately $-4.4$~kV, giving a drift field of approximately 350~V/cm. 

\begin{figure}
  \centering
  \adjincludegraphics[trim={{0.00000000\width} {0.14173263\height} {0.00000000\width} {0.00000000\height}},clip,width=0.8\columnwidth]{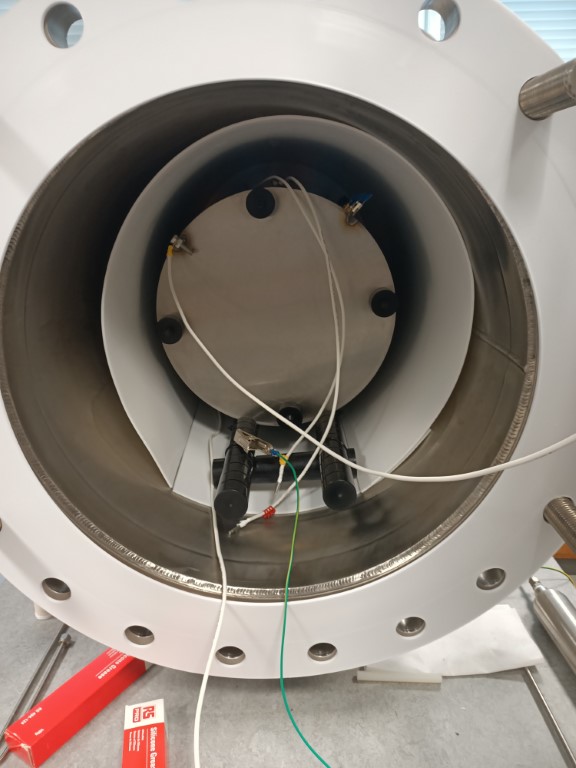}
  \caption{TPC installed in the pressure vessel, with the PTFE insulating sheet visible around the detector.}
  \label{fig:tpc-in-vessel}
\end{figure}

To reduce mechanical strain on the 50~M$\Omega$ resistor, its ground connection is made through a terminal block mounted in a holder spanning the two support rails. The resistor chain and terminal block holder can be seen in Fig.~\ref{fig:fc-full-assembly}. Figure~\ref{fig:tpc-in-vessel} shows the TPC inside the vessel, surrounded by a PTFE sheet to provide electrical insulation from the vessel.

\subsubsection{Amplification Stage}

Primary ionisation electrons drift towards the amplification stage, which consists of two THGEMs separated by a 3.5~mm gap. Each THGEM is 1~mm thick and has a circular active area 175~mm in diameter, with 0.5~mm-diameter holes at a pitch of 0.8~mm. The strong electric field within the holes accelerates the electrons, producing secondary ionisation and exciting the gas. These processes generate charge avalanches and secondary scintillation, forming the charge and optical signals.

\begin{figure}
  \centering
  \subfloat[Double-THGEM stack, with three M4 nylon bolts securing the holders together.\label{fig:thgem-double-stack}]{%
    \adjincludegraphics[trim={{0.09448842\width} {0.14173263\height} {0.09448842\width} {0.14173263\height}},clip,
      width=0.51\columnwidth]{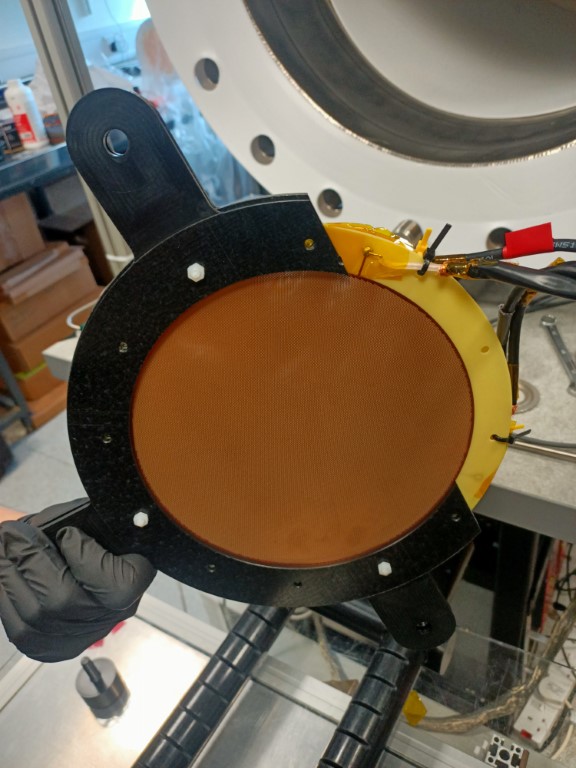}}
\hfill
  \subfloat[THGEMs and WLS mounted at the end of the field cage.\label{fig:wls-thgem-mounted}]{%
    \adjincludegraphics[trim={{0.07559073\width} {0.08503958\height} {0.14173263\width} {0.17716578\height}},clip,
      width=0.48\columnwidth]{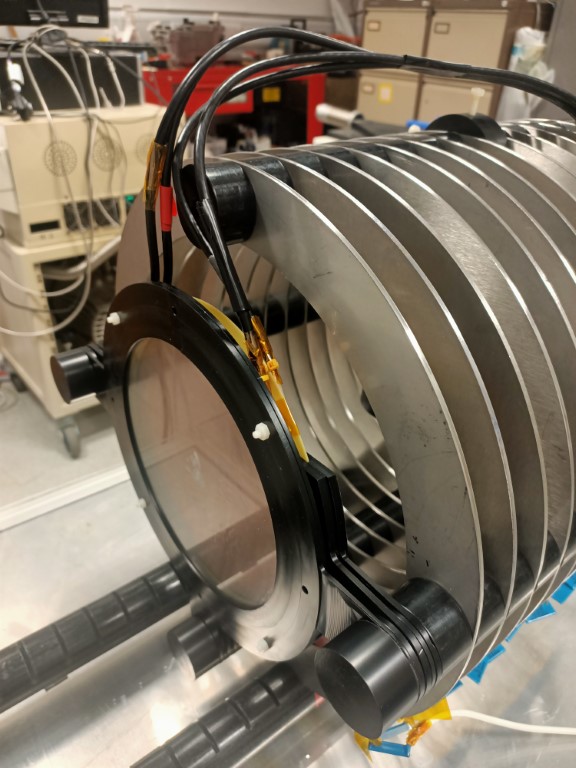}}

  \caption{Assembly and installation of the amplification stage.}
  \label{fig:wls-mounted}
\end{figure}

Each THGEM sits in a 1~mm-deep recess in a Delrin holder mounted at the end of the field cage. Three M4 nylon bolts secure the two holders together. Coaxial cables are soldered to the THGEM connection pads, with Kapton providing insulation and cable ties relieving the strain. The assembled stack is shown in Fig.~\ref{fig:wls-mounted}.

The four THGEM electrodes are numbered 1--4, from the one closest to the cathode to the furthest. They are powered by a four-channel CAEN N1471H 5.5~kV supply (Fig.~\ref{fig:HV_PS-diagram}). Although it can provide a voltage difference of up to 11~kV using oppositely biased outputs, the usable voltages in this work are limited by electrical breakdown in the THGEMs.

\subsection{Charge Readout}\label{subsubsec:thgems}

\begin{figure}
  \centering
  \includegraphics[width=\columnwidth]{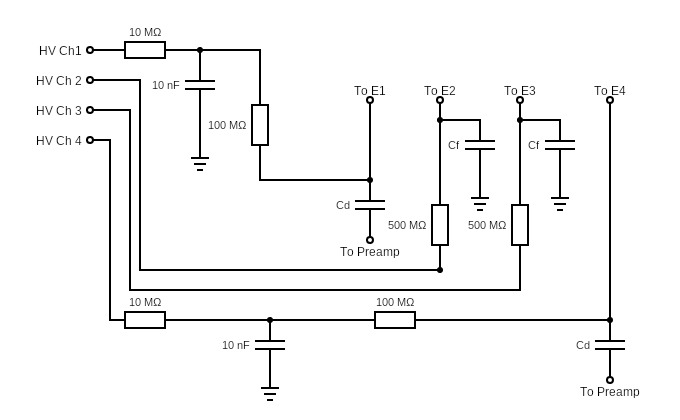}
  \caption{Circuit diagram of the resistor box  for the THGEM bias and charge readout.}
  \label{fig:new-resistor-box-circuit-diagram}
\end{figure}

Charge motion near the THGEM surfaces induces currents on the THGEM electrodes (Fig.~\ref{fig:HV_PS-diagram}). The induction signals from electrodes 1 and 4  are integrated by charge-sensitive preamplifiers and then digitised, forming the charge-readout signal. The charge-readout circuit is shown in Fig.~\ref{fig:new-resistor-box-circuit-diagram}.

Two Cremat CR-110 R2.2 charge-sensitive preamplifiers, each with a gain of 1.4~V/pC~\cite{cremat_cr110}, are mounted on CR-150-R6 CSP evaluation boards~\cite{CrematCR150R6}. A resistor box between the power supply and the detector readout circuit provides spark protection and a minimum load for the power supply. Bypass capacitors on electrodes 2 and 3 reduce crosstalk between the electrodes.

\subsubsection{Charge Data Acquisition}\label{subsubsec:charge-daq}

A PicoScope 5244D digitiser~\cite{pico5000d_datasheet} records the preamplifier outputs from electrodes 1 and 4 on two channels  at 125~MS/s. The PicoScope Software Development Kit (PicoSDK) provides the interface to the DAQ computer.

The trigger threshold is set to 50~mV. For each trigger, both channels record a waveform of 100~$\mu$s---50~$\mu$s before and 50~$\mu$s after the trigger. The full waveforms are saved in CSV format for analysis.


\subsection{Optical Readout}

The optical readout  system is based on a Timepix3 camera supplied by Amsterdam Scientific Instruments~\cite{ASI_TPX3Cam_2019}, together with an intensifier and lens. Light from the wavelength shifter (WLS) plane is focused by the lens onto the photocathode of the intensifier. Its output reaches the camera and is detected by its internal silicon sensor coupled to the Timepix3 readout chip.

\subsubsection{Wavelength Shifter}\label{subsubsec:wls}

Argon scintillates predominantly in the vacuum ultraviolet (VUV, near 128~nm) region. As the vessel viewport glass is opaque to such VUV light~\cite{Martinez2025ARIADNEPlus}, the light must be shifted to longer wavelengths so that it can pass through the glass. We use polyethylene naphthalate (PEN)~\cite{Direct_comparison_of_PEN_and_TPB_wavelength_shifters_in_a_liquid_argon_detector,Development_of_wavelength-shifting_PEN_foils_for_next_generation_experiments} as the WLS, with an emission peak near 420~nm. The assembly consists of a 2-mm-thick, 200-mm-diameter heat-strengthened borosilicate-glass disc as the base. A PEN sheet is glued to the base using a two-part epoxy resin with good optical properties. The assembly is secured to a 2~mm-deep recess in a Delrin holder (Fig.~\ref{fig:wls-thgem-mounted}). 

\subsubsection{Camera System}

The lens, intensifier, and the camera are connected through two C-mount connections. 

The lens sits approximately 70~cm from the WLS. It has a maximum aperture of f/0.95, a focal length of 50~mm, and a minimum focusing distance of 600~mm. The model used, Navitar DO-5095, is suitable for low-light imaging~\cite{NavitarHighSpeed}. 

The intensifier is a Photonis Cricket 2 image intensifier with a high-quantum-efficiency UV photocathode sensitive between 200 and 450~nm, covering the emission peak of the WLS. Photoelectrons from the photocathode are multiplied by two microchannel plates with a gain of approximately $10^{6}$, before reaching a P47 phosphor screen~\cite{Optical_Readout_of_the_ARIADNE_LArTPC_using_a_Timepix3_based_Camera}. This screen consists of cerium-doped yttrium silicate ($\mathrm{Y_2SiO_5:Ce}$) that emits between 370 and 480~nm with a decay time of approximately 100~ns from 90\% to 10\% of its initial intensity~\cite{PhotonisPhosphorScreens}. 

The camera contains a  300~$\mu$m-thick silicon sensor, bump-bonded to a Timepix3 chip. The chip has $256 \times 256$ pixels at a pitch of 55~$\mu$m, giving a sensitive area of 1.98~cm$^2$~\cite{Gromov2012Timepix3}.

\begin{figure}
    \centering
    \adjincludegraphics[trim={{0.33307167\width} {0.23622105\height} {0.14173263\width} {0.37795367\height}},clip,width=0.9\columnwidth]{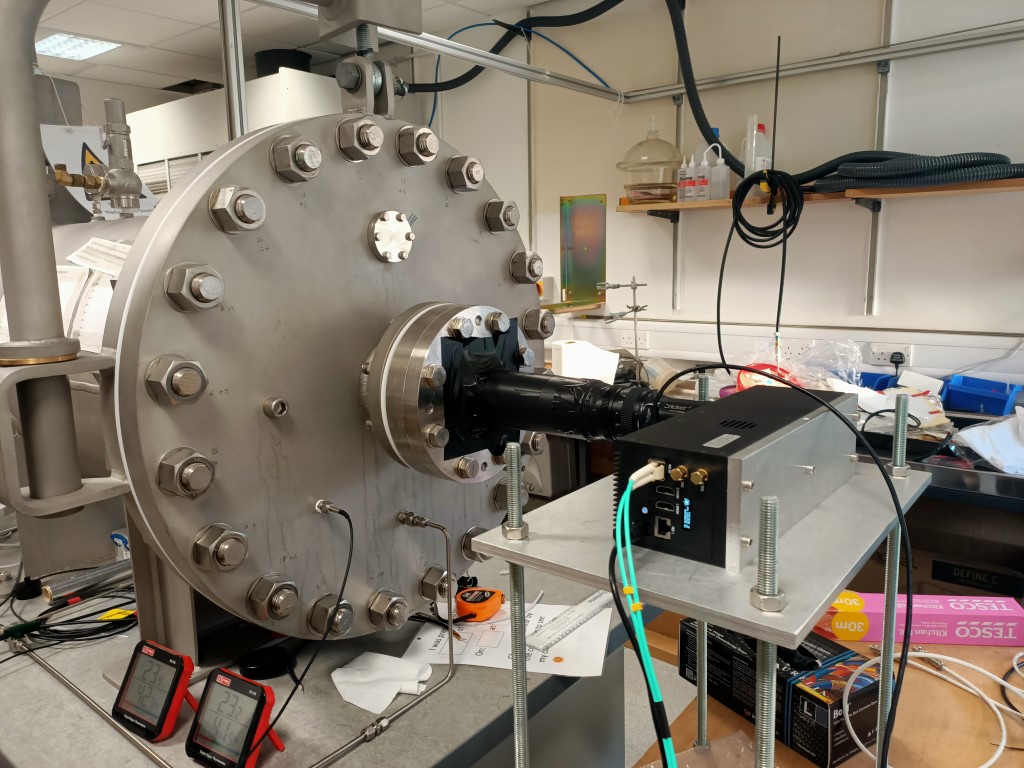}
    \caption{The Timepix3-based camera system, including the light shield, lens, intensifier, and the camera. The light shield is secured to the viewport with electrical tape.}
    \label{fig:tpx-setup}
\end{figure}

The camera is mounted on an elevated metal plate secured to a table with threaded bolts. Its height and angle can be adjusted using a spirit level and subsequent inspection of the camera images. A 3D-printed light shield also helps to improve the alignment. The system is shown in Fig.~\ref{fig:tpx-setup}.

\subsubsection{Optical Data Acquisition}\label{subsubsec:optical-daq}

A SPIDR board inside the camera housing transmits the Timepix3 data to the DAQ computer through a 10~Gbps Ethernet connection. The data are saved in the raw binary \texttt{.tpx3} format and subsequently converted to \texttt{.hdf5} files using the LUNA package~\cite{ASI_LUNA}.

A pixel signal above the detection threshold is registered as a hit. For each hit, the pixel coordinates, time of arrival (ToA), and time over threshold (ToT) are recorded~\cite{Timepix3_a_65K_channel_hybrid_pixel_readout_chip_with_simultaneous_ToA/ToT_and_sparse_readout}. The readout  records only hits (zero-suppressed). 

Timing is provided by a 40~MHz clock and a 640~MHz voltage-controlled oscillator (VCO)~\cite{Timepix3_a_65K_channel_hybrid_pixel_readout_chip_with_simultaneous_ToA/ToT_and_sparse_readout,Timepix3_first_measurements_and_characterization_of_a_hybridpixel_detector_working_in_event_driven_mode,Design_of_the_analog_frontend_for_the_Timepix3_and_Smallpix_hybrid_pixel_detectors_in_130_nm_CMOS_technology}. The 40~MHz clock provides the coarse ToA measurement. When a hit is registered, the VCO measures the interval to the next rising edge of the coarse clock, providing a fine ToA with a bin width of 1.56~ns ($=1/640$~MHz)~\cite{Gromov2012Timepix3,Timepix3_a_65K_channel_hybrid_pixel_readout_chip_with_simultaneous_ToA/ToT_and_sparse_readout}. As in a general TPC, the ToA can be converted to a drift coordinate using  the drift velocity, thereby providing 3D tracking information~\cite{Optical_Readout_of_the_ARIADNE_LArTPC_using_a_Timepix3_based_Camera}.

The ToT measures the duration of a hit, using the 40~MHz clock with a bin width of 25~ns ($=1/40$~MHz). It is related to the charge collected by the pixel and therefore represents the optical response.

\section{Detector Commissioning}
\label{sec:3-procedure}

As each subsystem was integrated into the platform, tests were performed to verify  detector performance, including leak tests of feedthroughs and valves, gas-loss rate measurements, electronic-noise mitigation, and cathode and transfer-field scans. This paper focuses on  tracking observation and optical response measurement; other commissioning studies can be found in Ref.~\cite{SnapeThesis}.

\subsection{Radioactive Source}\label{subsubsec:sourceholder}

\begin{figure}
  \centering
  \subfloat[Source holder with the protective gate.]{\adjincludegraphics[trim={{0.29291410\width} {0.29055189\height} {0.23622105\width} {0.33307167\height}},clip,width=0.56\columnwidth]{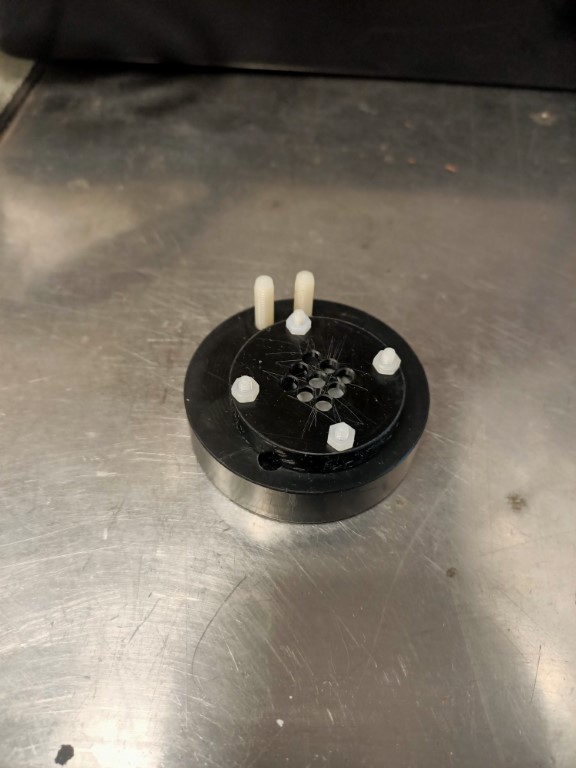}\label{fig:source-tray}}
  \hfill
  \subfloat[Source holder inside the field cage.]{\adjincludegraphics[trim={{0.18897684\width} {0.35433157\height} {0.23622105\width} {0.03543316\height}},clip,width=0.42\columnwidth]{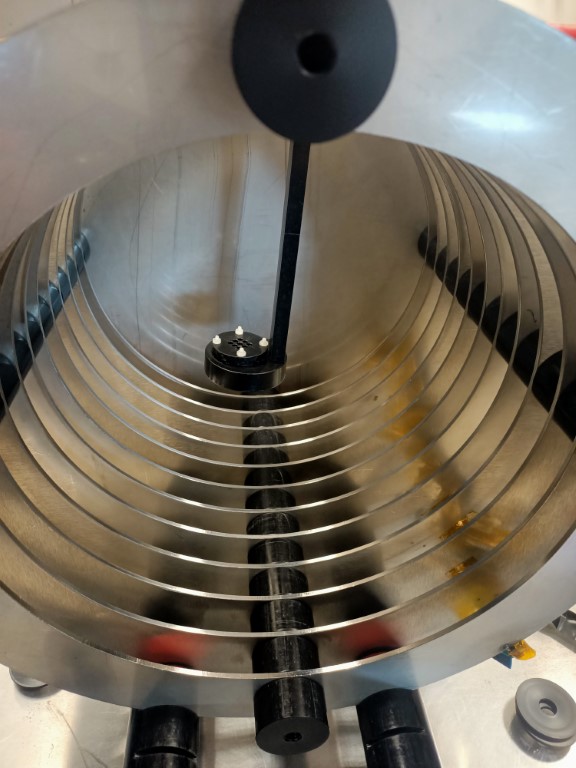}\label{fig:source-in-field-cage}}
  \caption{The radioactive source holder (a) before installation and (b) mounted inside the TPC.
  }\label{fig:source-mounted}
\end{figure}

In this work, we use a $^{241}$Am alpha source for  commissioning. The alpha particles have a kinetic energy of 5.4~MeV, and the source has a  nominal activity of 3~kBq. The source housing is 25~mm (16~mm active region) in diameter and 3~mm thick, with a minimum operating pressure of 0.25~bar. 

A source holder is built to secure the source in a recessed tray beneath a protective gate (Fig.~\ref{fig:source-mounted}). The gate has a group of $\diameter$4~mm holes that allow the alpha particles to pass through. The holder is suspended from a Delrin hook attached to a spacer of the field cage. 

\subsection{$\alpha$-Particle and Cosmic-Ray Tracks}

\begin{figure*}
 \centering
    \begin{subfigure}[t]{0.49\linewidth}
        \centering
        \includegraphics[width=\linewidth]{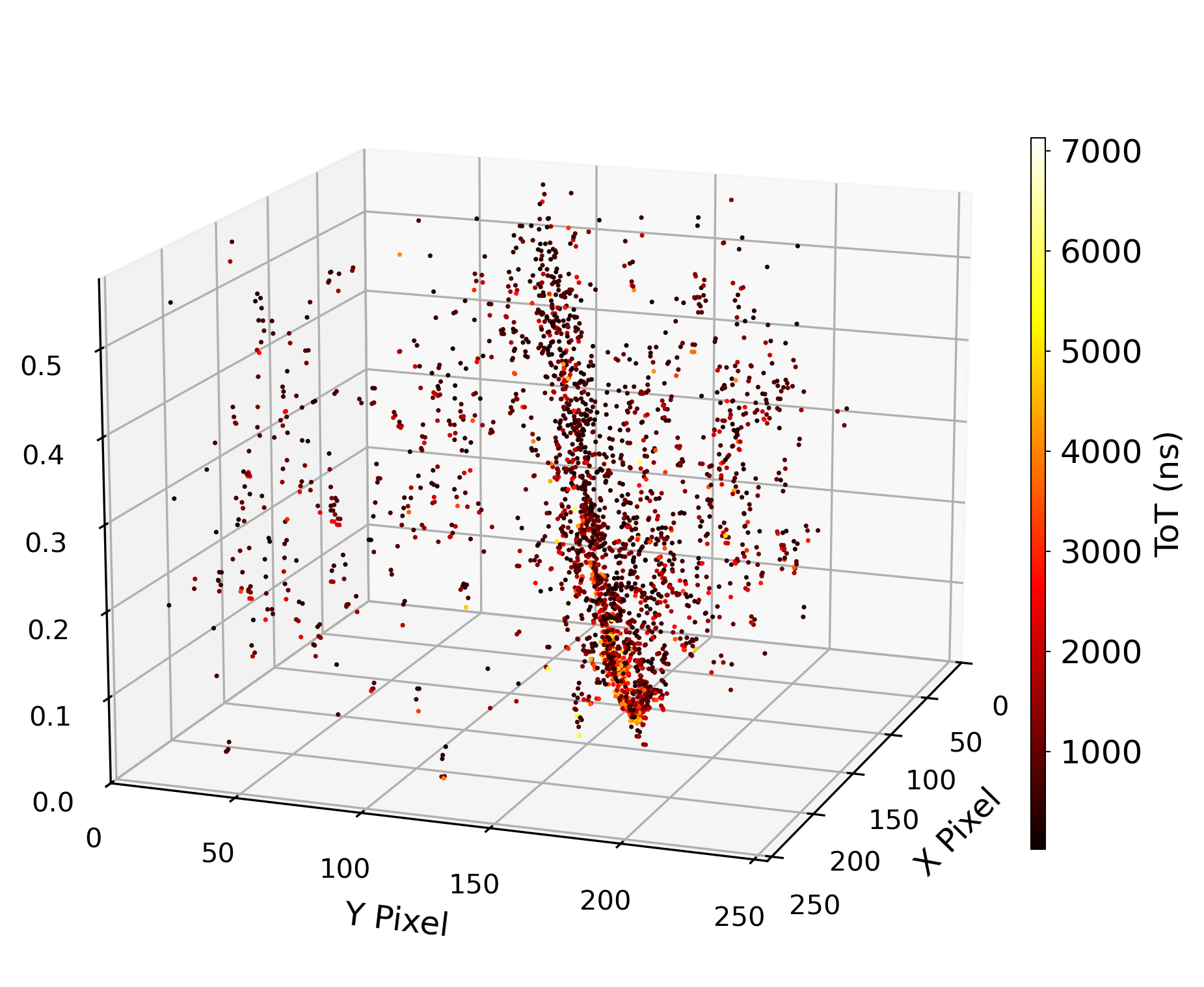}
        \caption{}
        \label{alpha_track_3D}
    \end{subfigure}\hfill
    \begin{subfigure}[t]{0.49\linewidth}
        \centering
        \includegraphics[width=\linewidth]{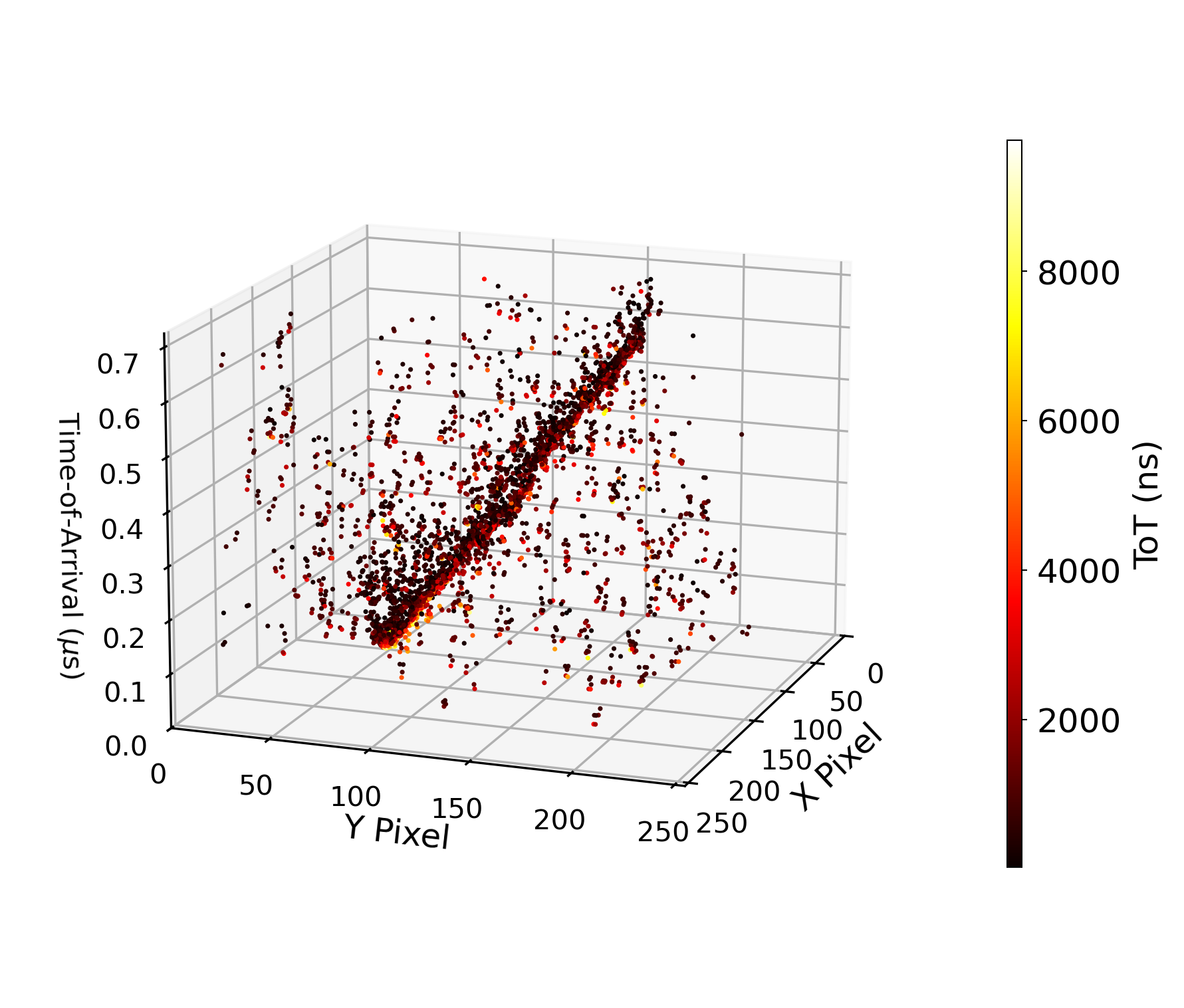}
        \caption{}
        \label{cosmic_track_3D}
    \end{subfigure}
     \medskip
    \begin{subfigure}[t]{0.49\linewidth}
        \centering
        \includegraphics[width=\linewidth]{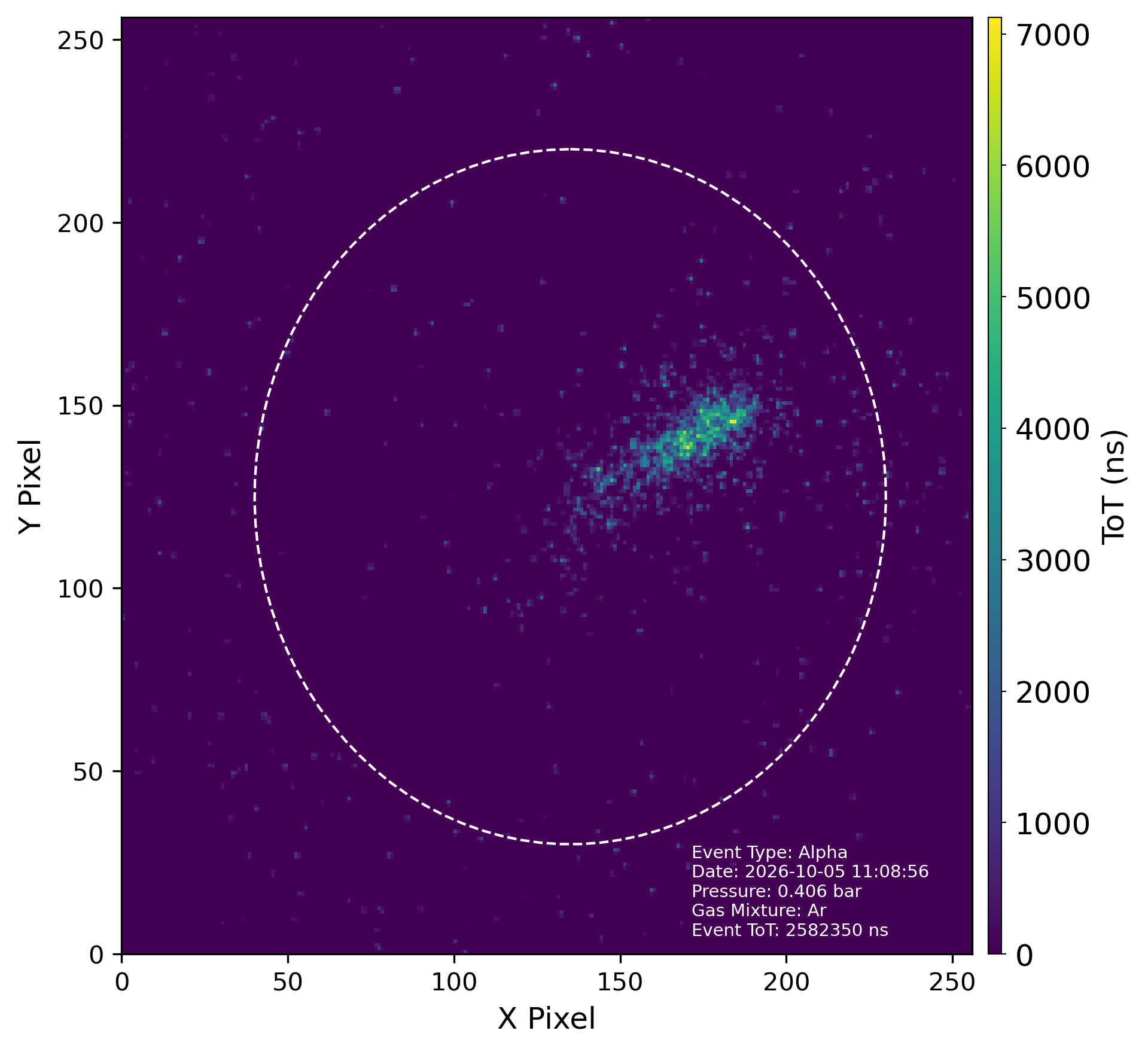}
        \caption{}
        \label{alpha_track_2D}
    \end{subfigure}\hfill
    \begin{subfigure}[t]{0.49\linewidth}
        \centering
        \includegraphics[width=\linewidth]{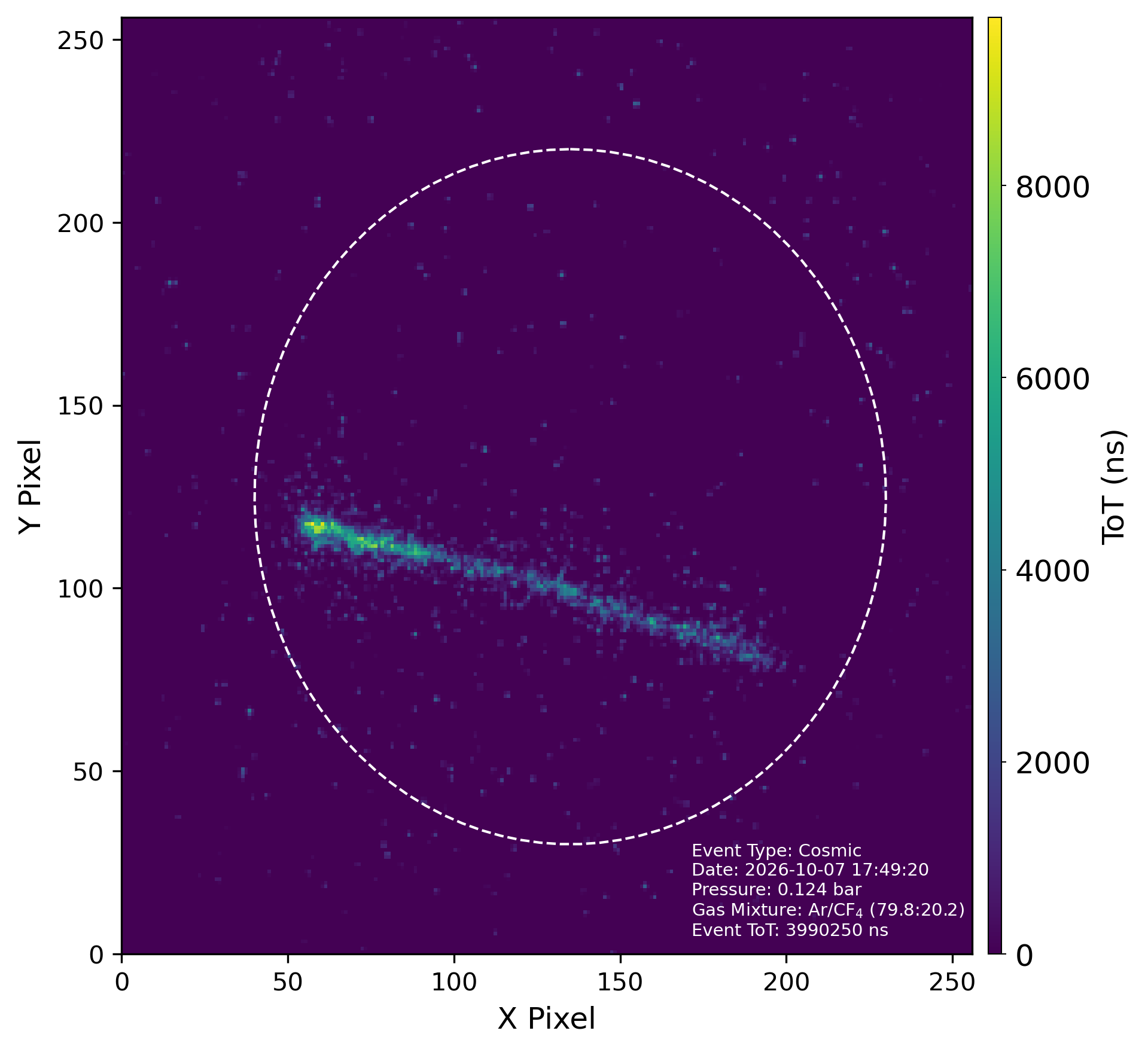}
        \caption{}
        \label{cosmic_track_2D}
    \end{subfigure}
    \caption{
    Event displays of (a) an alpha particle in argon at 0.4~bar and (b) a cosmic ray in an 80:20 argon--CF$_4$ mixture at 0.1~bar, both in pixel coordinates and ToA. (c) and (d) show their projections onto the image plane. All colour scales indicate ToT.
    }
    \label{tracks}
\end{figure*}

We report the observation of alpha-particle tracks from a $^{241}$Am source in argon at 0.4~bar, and of cosmic-ray tracks in an 80:20 argon--CF$_4$ mixture at 0.1~bar with the source removed. Figure~\ref{tracks} shows examples of these events. The pixel coordinates are the spatial projection onto the THGEM image plane, while the ToA could be converted into a spatial coordinate using the drift velocity in the future, and the ToT is the optical response. Detailed track reconstruction and physics analysis will be reported in a subsequent paper.

\subsection{Optical Response Measurement}

\begin{figure} 
  \centering
  \includegraphics[width=0.8\columnwidth]{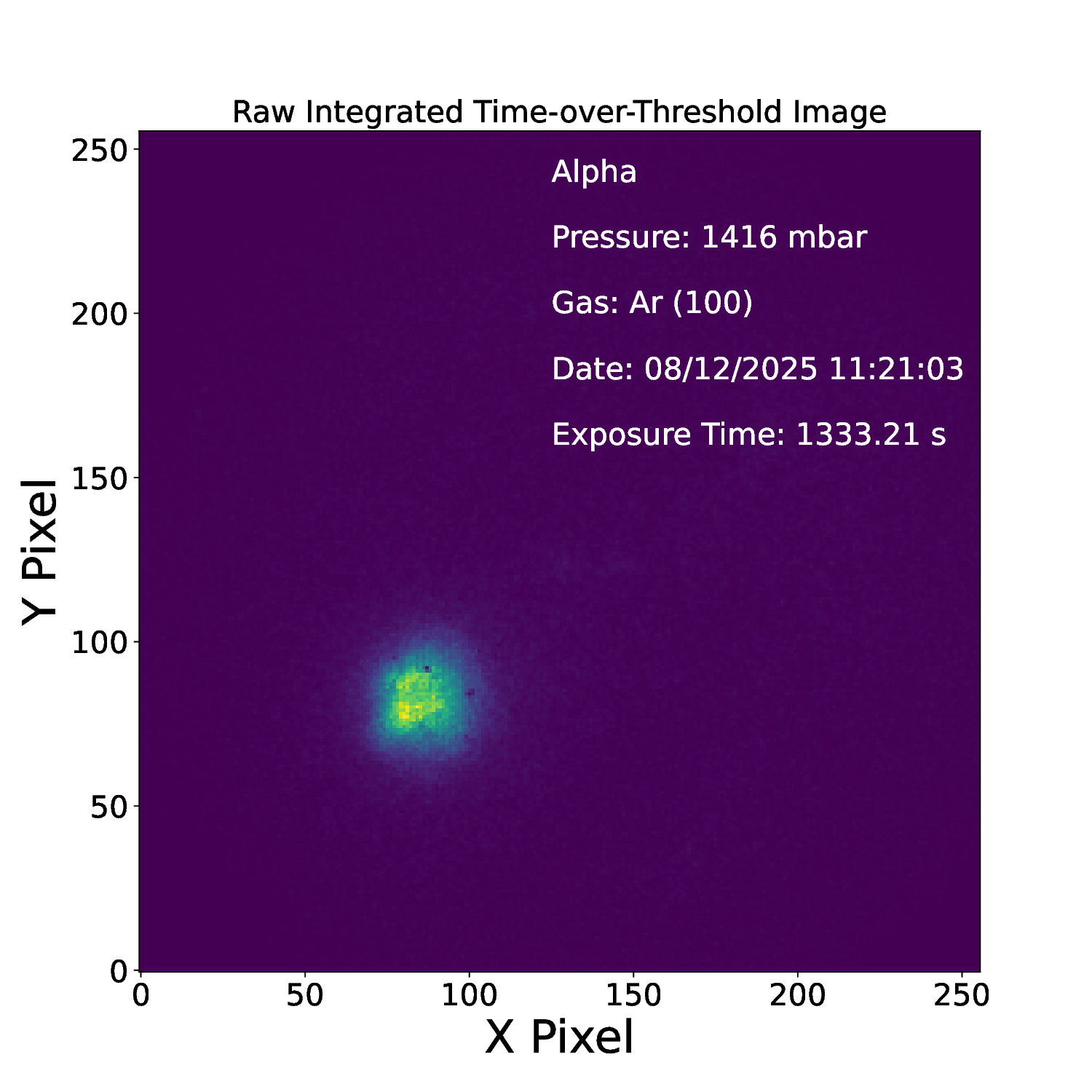}
  \caption{Time-integrated optical image of alpha particles from the $^{241}$Am source. }
  \label{fig:img-sideband-varx}
\end{figure}

The optical response measurement uses data from the $^{241}$Am source  accumulated over an exposure of approximately 20~minutes for each given running condition. Instead of images of tracks, we have an integrated image of a diffuse spot formed by overlaid alpha-particle tracks (Fig.~\ref{fig:img-sideband-varx}). 

Summing the pixel-by-pixel ToT values in the time-integrated image gives the total optical signal. Dividing it by the exposure time gives the normalised optical signal $S_{\mathrm{O}}^{\mathrm{norm}}$,  in units of ms/s. Because the light originates from the secondary ionisation of electrons, this optical response inherently includes the charge gain. Measuring the optical gain requires charge readout and charge gain calibration. 

The optical response is characterised as a function of the gas mixture and the total reduced voltage across the two THGEMs,
\begin{equation}
    \frac{V_{43+21}}{p}
    \equiv \frac{(V_4-V_3)+(V_2-V_1)}{p},
\end{equation}
where $V_1$ and $V_2$ are the voltages applied  to THGEM1, $V_3$ and $V_4$ are those applied to THGEM2, as shown in Fig.~\ref{fig:HV_PS-diagram}, and $p$ is the absolute gas pressure. Three commissioning runs with varying conditions are reported in this work:
\begin{itemize}
    \item \textit{Voltage run:} $V_2$ and $V_3$ are held at $-100$~and 0~V, respectively. $V_1$ and $V_4$ are varied simultaneously from $-1000$ and $1100$~V to $-1400$ and $1500$~V, respectively, each changing in steps of 50~V.  
    \item \textit{Pressure run:} The pressure is increased from 1 to 1.7~bar in steps of 0.1~bar, and then from 2 to 7.5~bar in steps of 0.5~bar. Results at higher pressures are not reported, as the signal is too dim to analyse.
    \item \textit{Methane run:} Methane is added to argon in increasing fractions up to 10\% in steps of approximately 0.5\%.
\end{itemize}
During each run, all other conditions are held constant. 

\subsubsection{Optical Data Analysis}\label{sec:analysis}

\begin{figure}
  \centering
  \begin{subfigure}[t]{0.49\columnwidth}
    \centering
    \includegraphics[width=\linewidth]{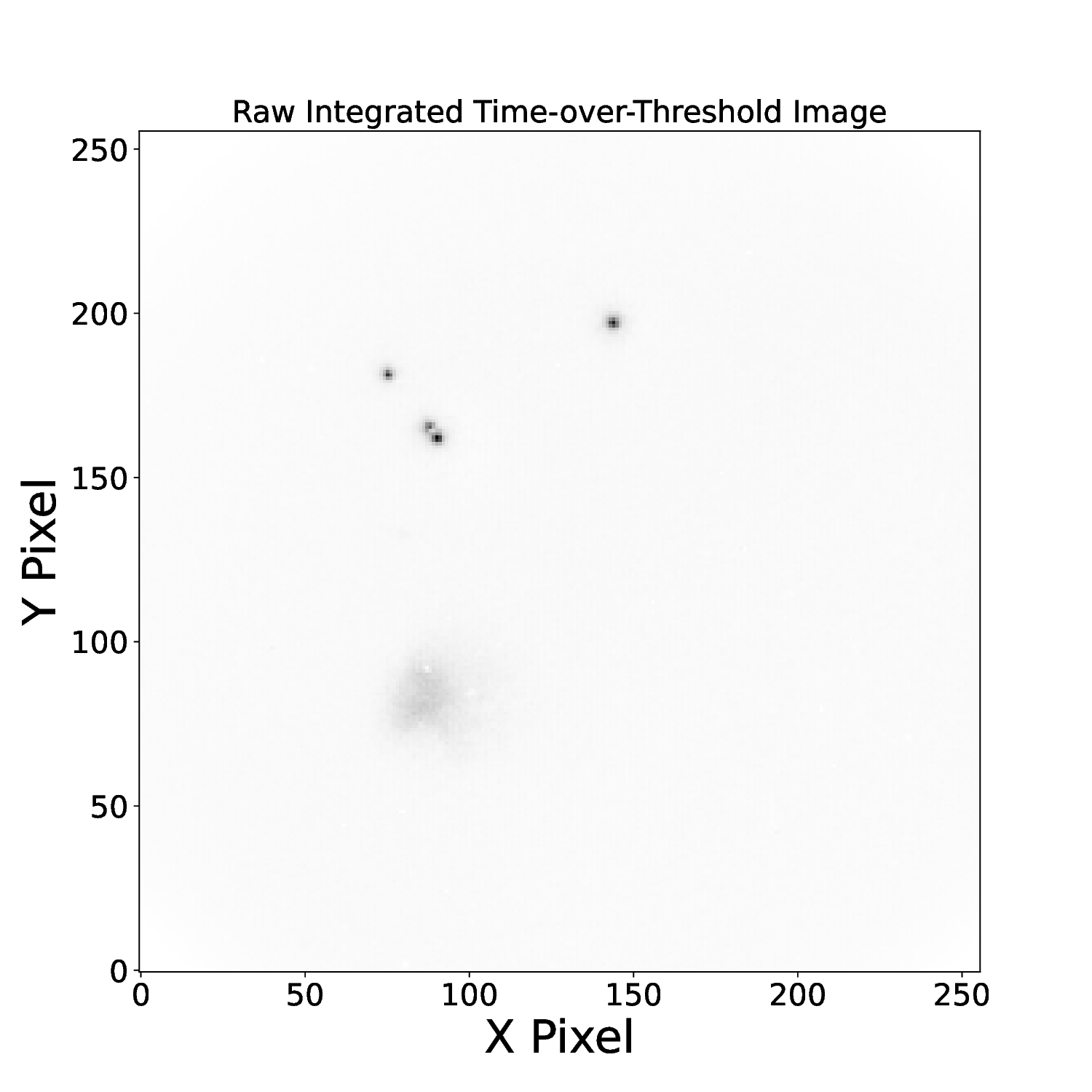}
    \caption{Before spark removal.}
    \label{fig:spark-img}
  \end{subfigure}
  \hfill
  \begin{subfigure}[t]{0.49\columnwidth}
    \centering
    \includegraphics[width=\linewidth]{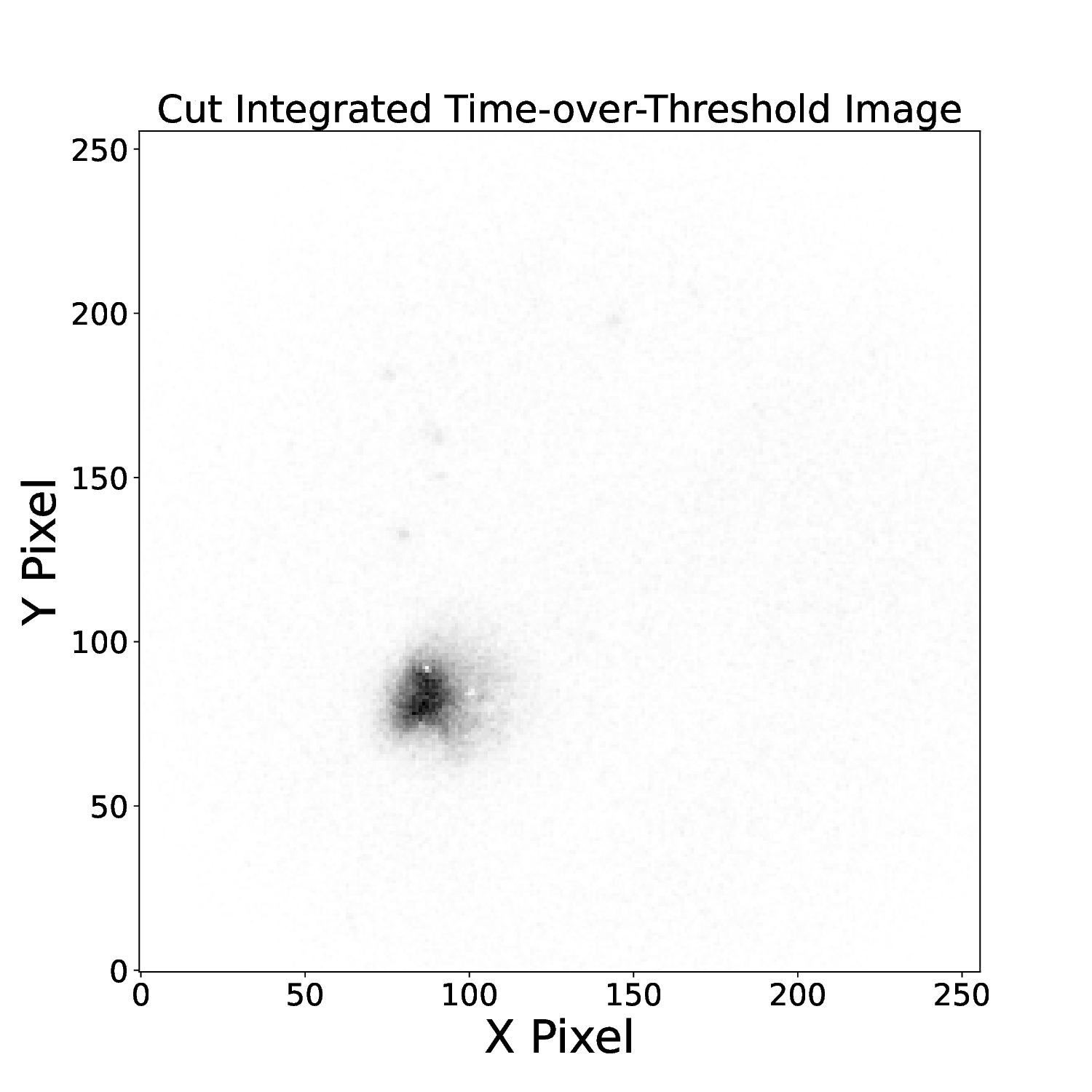}
    \caption{After spark removal.}
    \label{fig:cut-methane-images}
  \end{subfigure}
  \caption{
  Optical images before and after spark removal. Data taken in an Ar--CH$_4$ (97:3) mixture at 1210~mbar on 11 December 2025, with a total exposure of 1333.19~s.}
  \label{fig:spark-removal}
\end{figure}

\textit{Spark removal}: Electric-field instabilities in the THGEMs can cause sparks, leaving bright spots in the optical image. Although such events are infrequent, they need to be removed from the data set. The procedure is to discard those time frames that contain the largest integrated ToT signals (the top 1\% in this work). It is an effective procedure, as is shown in Fig.~\ref{fig:spark-removal}. 

\begin{figure}
  \centering
  \subfloat[]{\includegraphics[width=0.49\columnwidth]{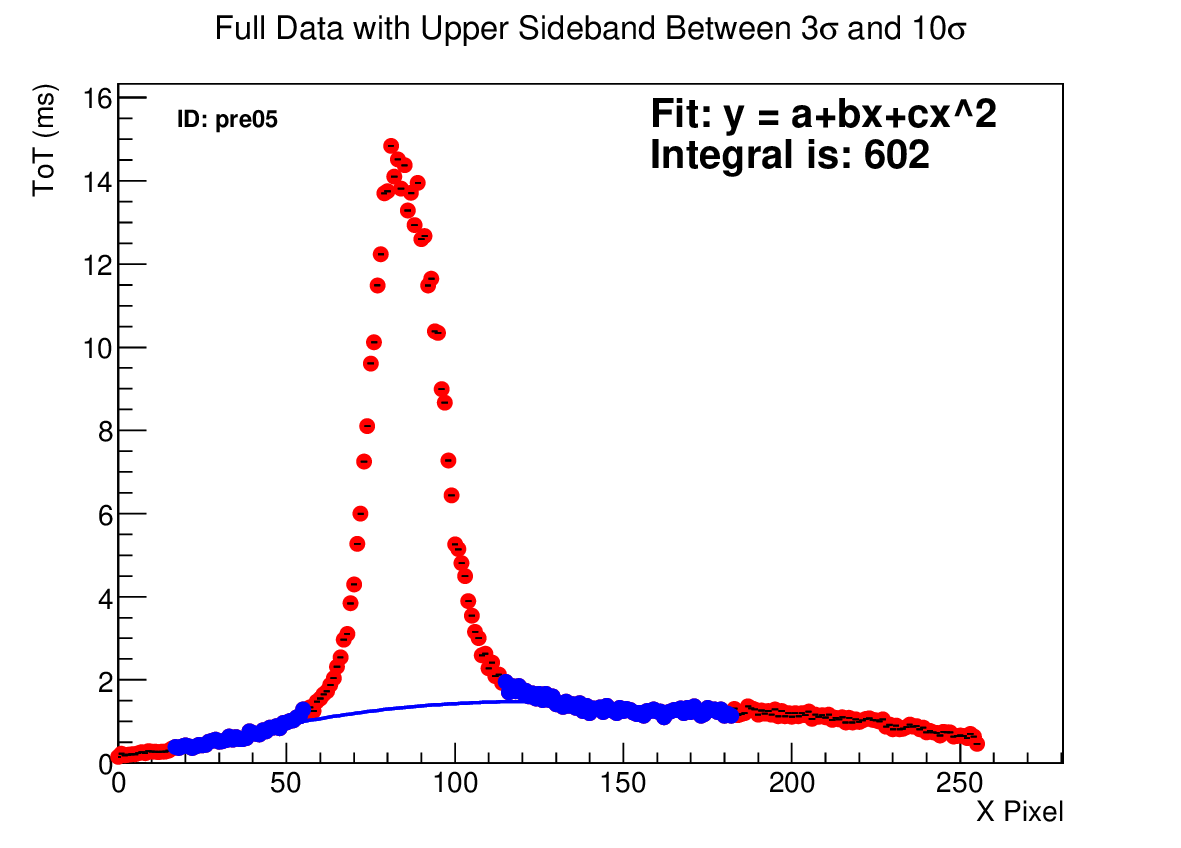}\label{fig:1416-fit}}
  \hfill
  \subfloat[]{\includegraphics[width=0.49\columnwidth]{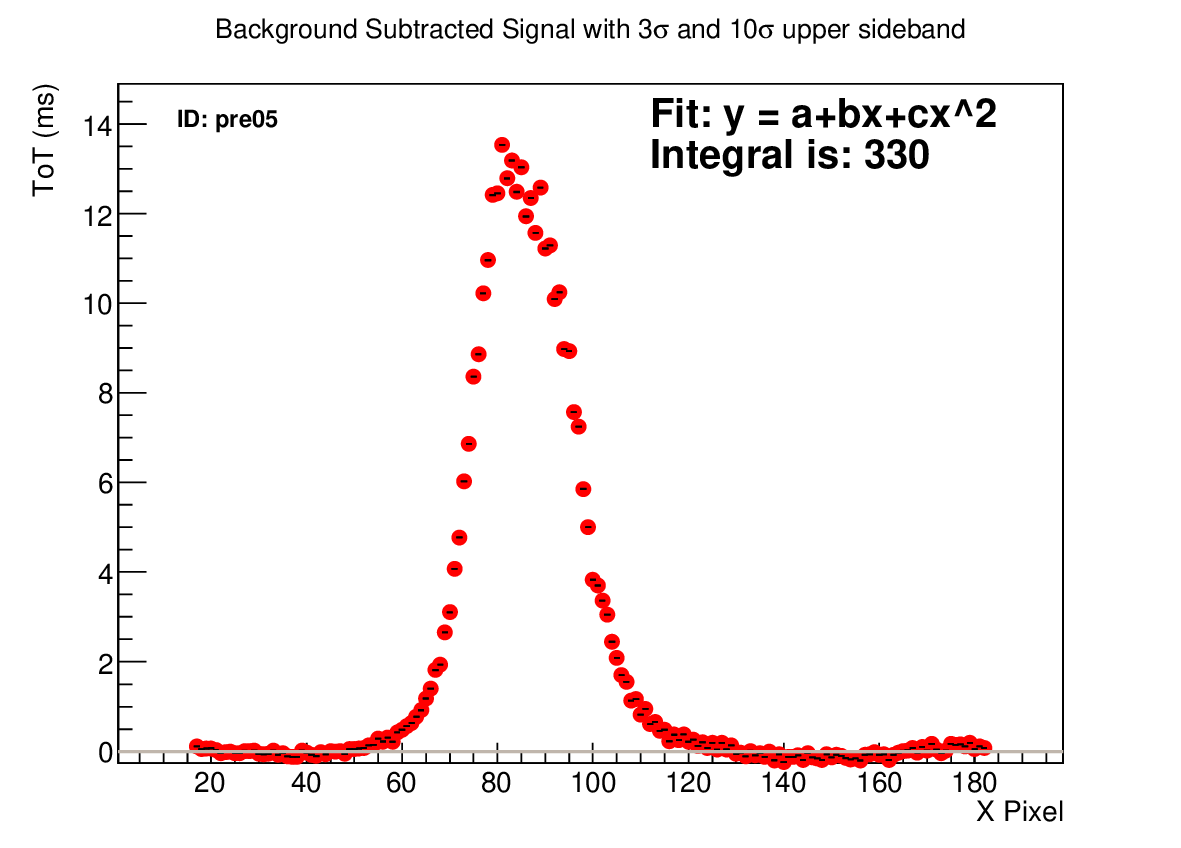}\label{fig:1416-sig}}
  \caption{Exemplary sideband fit to data and the resulting background-subtracted distribution.}
  \label{fig:opt-sideband-fitting}
\end{figure}

\textit{Sideband fit}: The time-integrated image, recording ToT as a function of the X and Y pixel coordinates over the total exposure, contains both the signal originating from primary gas ionisation and backgrounds. To first order, it is a 2D Gaussian peak on top of a flat background. The exact shapes of the peak and background vary with experimental conditions. Therefore, it is important to model the background with enough flexibility while determining the signal size in a way that is independent of the peak shape. The procedure used in this work is to 
\begin{enumerate}
    \item define a sideband region sufficiently far from the peak in the 2D image,
    \item project the 2D image onto one axis to form a 1D distribution, where the sideband is fitted with a polynomial to estimate the background under the peak, and
    \item subtract the fitted background from the peak region and determine the size of the peak bin by bin.
\end{enumerate}
Figure~\ref{fig:opt-sideband-fitting} shows an example of the sideband fit and the resulting signal.

\textit{Systematic uncertainty}: The background-subtracted ToT signal is then integrated along the projected pixel axis to obtain $S_{\mathrm{O}}$. The exposure time, determined from the ToA, is used to normalise this signal, giving the normalised optical signal $S_{\mathrm{O}}^{\mathrm{norm}}$.

\begin{figure}
  \centering
  \includegraphics[width=0.8\columnwidth]{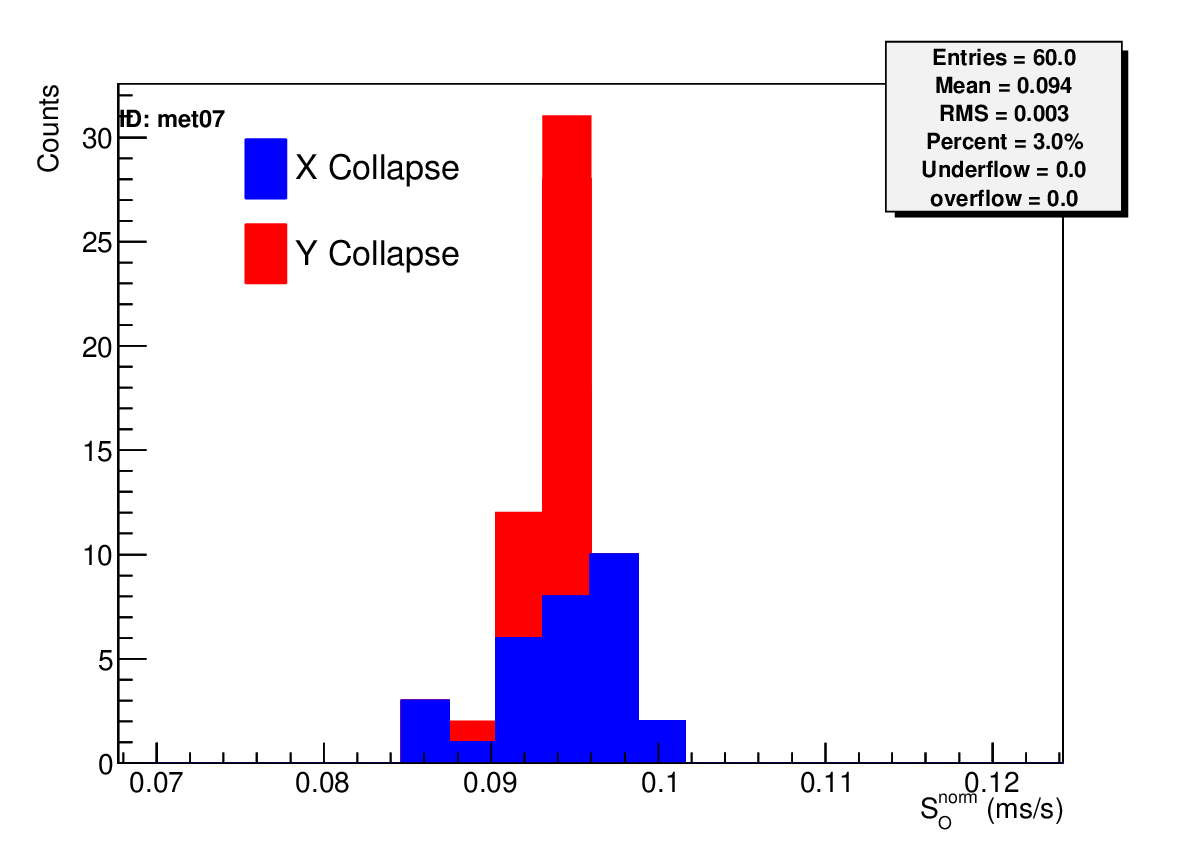}
  \caption{Distribution of the normalised optical signal $S_{\mathrm{O}}^{\mathrm{norm}}$ by varying the projection axis, sideband selection, and the background polynomial order.} 
  \label{fig:sig-hists}
\end{figure}

The fitted background shape can change significantly with the choice of  sideband  range and  fitting function. Therefore, we use combinations of 10 different sideband ranges and polynomials of 3 different orders. The image projection can also vary the background shape because the TimePix3 chip readout proceeds sequentially in one direction. In total, 60 variations are considered (Fig.~\ref{fig:sig-hists}). The mean  of the resulting $S_{\mathrm{O}}^{\mathrm{norm}}$ distribution is taken as the measured value with its RMS as the systematic uncertainty.

\subsubsection{Results}\label{sec:4-res}

\begin{figure}
    \centering
    \adjincludegraphics[trim={{0.00000000\width} {0.00000000\height} {0.00000000\width} {0.06964748\height}},width=0.8\columnwidth,clip]{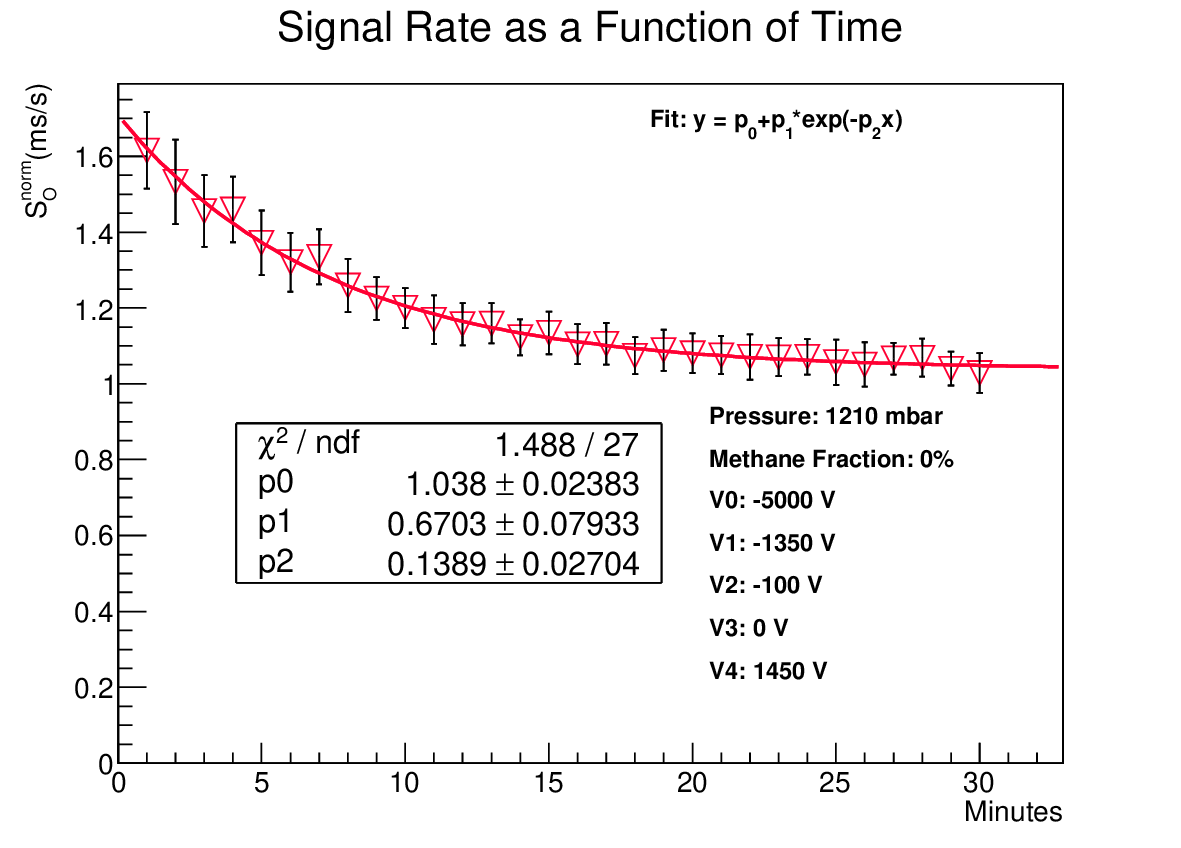}
    \caption{$S_{\mathrm{O}}^{\mathrm{norm}}$ as a function of time of voltage application. A fit is presented to describe the charging-up effect.}
    \label{fig:charge-up-graph}
\end{figure}

\textit{Charging-up effect:} When an electric field is established in a THGEM, it takes time to stabilise due to charge accumulation on the substrate surfaces~\cite{Bressler2023THGEM,Breskin2009THGEM}. This effect is measured and shown in Fig.~\ref{fig:charge-up-graph}. The normalised optical signal $S_{\mathrm{O}}^{\mathrm{norm}}$ decreases exponentially: 
\begin{equation}
    S_{\mathrm{O}}^{\mathrm{norm}}(t) = p_0 + p_1 \mathrm{e}^{-p_2 t},\nonumber
\end{equation}
with a characteristic time of $1/p_2 = 7.2 \pm 1.4$~minutes. It stabilises at $p_0/(p_0+p_1)=61\%$ of its initial value. This behaviour is consistent with the measurement previously reported in Ref.~\cite{A_Novel_Manufacturing_Process_for_Glass_THGEMs_and_First_Characterisation_in_an_Optical_Gaseous_Argon_TPC}. As a result, we bias the THGEMs typically for one hour before taking data in the commissioning runs.

\begin{figure}
  \centering
  \adjincludegraphics[trim={{0.00000000\width} {0.00000000\height} {0.00000000\width} {0.08357698\height}},width=\columnwidth,clip]{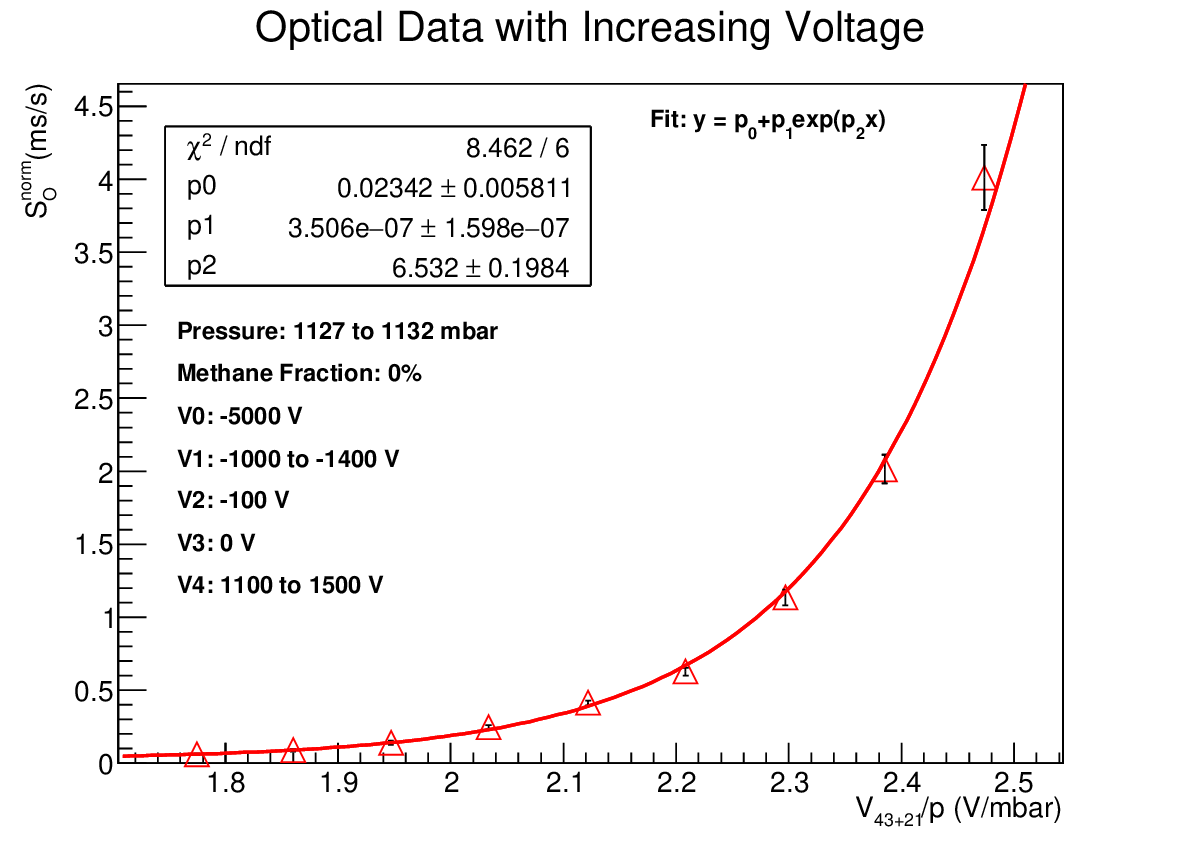}
  \caption{$S_{\mathrm{O}}^{\mathrm{norm}}$ as a function of the total reduced voltage across the two THGEMs $V_{43+21}/p$ in the \textit{voltage run}. An empirical fit to the data is shown.}
  \label{fig:volt-opt-lin}
\end{figure}

\textit{Voltage run}: $S_{\mathrm{O}}^{\mathrm{norm}}$ is shown in Fig.~\ref{fig:volt-opt-lin} to increase exponentially with  the total reduced voltage across the two THGEMs, $V_{43+21}/p$. It reaches approximately 4~ms/s at 2.5~V/mbar, corresponding to an increase by a factor of approximately 70 ($\simeq4/0.06$) compared to the lowest data point.

\begin{figure}
  \centering
  \adjincludegraphics[trim={{0.00000000\width} {0.00000000\height} {0.00000000\width} {0.08357698\height}},width=\columnwidth,clip]{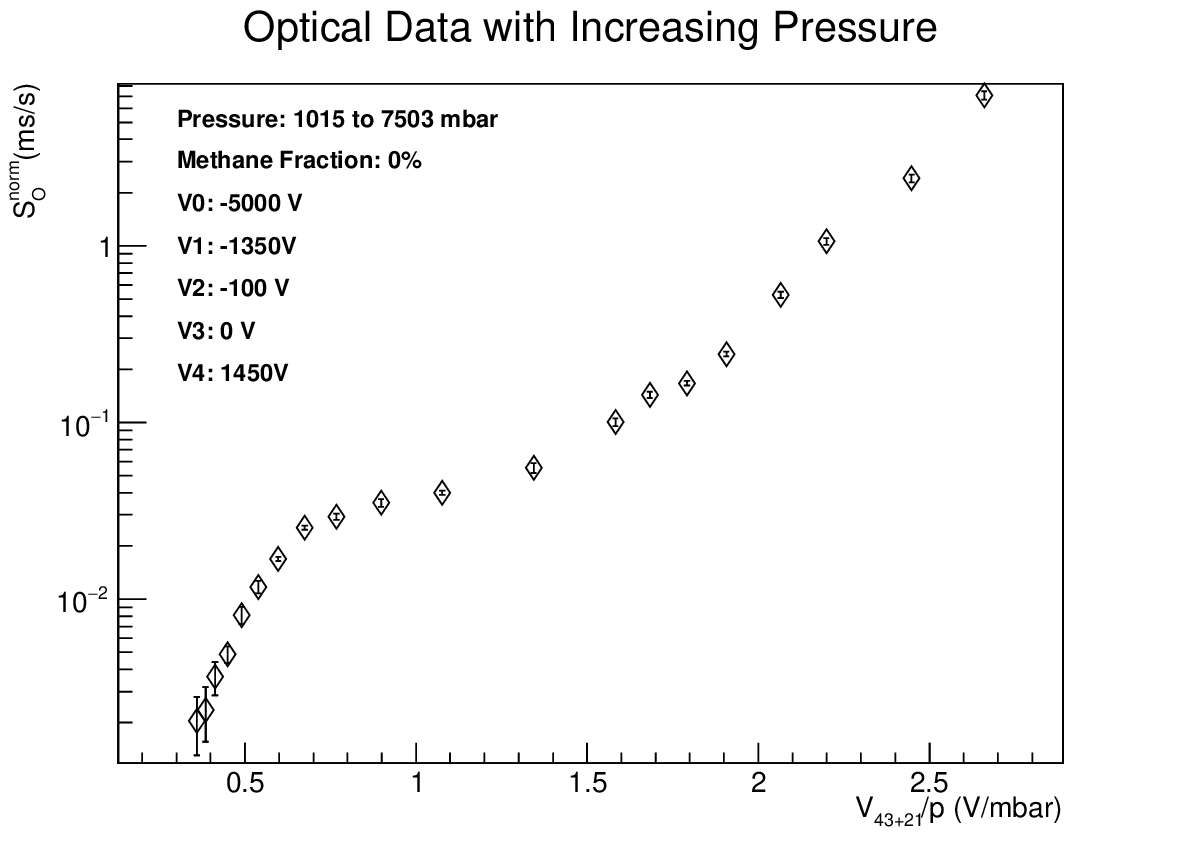}
  \caption{$S_{\mathrm{O}}^{\mathrm{norm}}$ as a function of $V_{43+21}/p$ in the \textit{pressure run}.}
  \label{fig:pres-opt-log}
\end{figure}

\textit{Pressure run}: $S_{\mathrm{O}}^{\mathrm{norm}}$ is shown in Fig.~\ref{fig:pres-opt-log} to decrease with increasing pressure. From 1 to 7.5~bar, the signal is reduced by a factor of approximately 3500: As $V_{43+21}/p$ drops from  2.7 to 1.5~V/mbar, the signal decreases approximately exponentially; between 1.5 and 0.7~V/mbar, the signal reaches a slow plateau; below 0.7~V/mbar, it falls rapidly as the pressure increases further. This change in the fall-off tendency could be due to a change in the relative contributions of VUV electroluminescence and neutral bremsstrahlung~\cite{Neutral_bremsstrahlung_and_excimer_electroluminescence_in-noble_gases_and_its_relevance_to_two_phase_dark_matter_detectors,Revealing_neutral_bremsstrahlung_in_two-phase_argon_electroluminescence}. However, this measurement alone does not distinguish between these emission mechanisms. 

\begin{figure}
  \centering
  \adjincludegraphics[trim={{0.00000000\width} {0.00000000\height} {0.00000000\width} {0.08357698\height}},width=\columnwidth,clip]{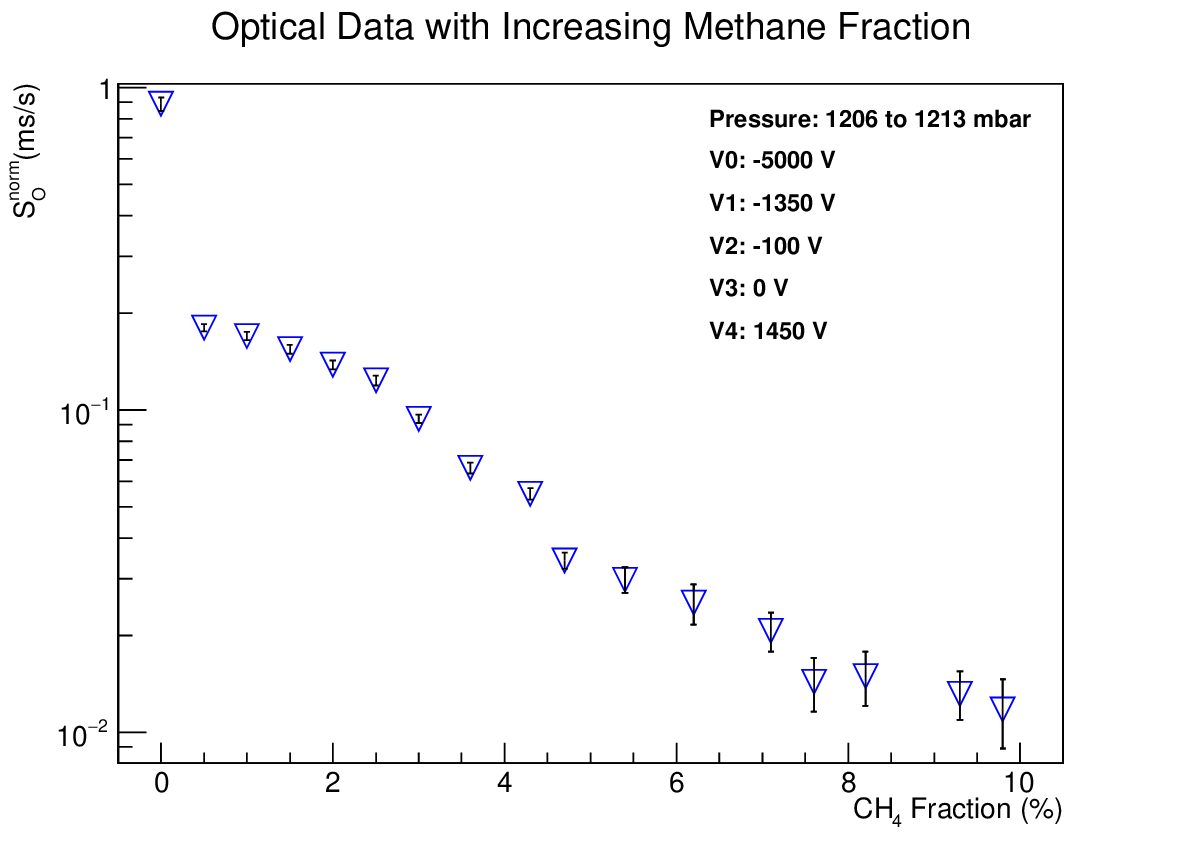}
  \caption{$S_{\mathrm{O}}^{\mathrm{norm}}$ as a function of the methane fraction in the \textit{methane run}. }
  \label{fig:met-opt-lin}
\end{figure}

\textit{Methane run}: $S_{\mathrm{O}}^{\mathrm{norm}}$ is shown in Fig.~\ref{fig:met-opt-lin} to drop rapidly when methane is added to argon: a signal reduction of a factor of approximately 5 is seen when  0.5\% methane is added; a more gradual reduction---reaching nearly 100 fold---is observed as the methane fraction increases to approximately 10\%. The signal reduction may result from a suppression of VUV photons through absorption and non-radiative de-excitation of argon excimers. Similar studies using methane-doped liquid argon have been reported~\cite{Observation-of-primary-scintillations-in-the-visible-range-in-liquid-argon-doped-with-methane}.

\begin{figure}
    \centering
    \adjincludegraphics[trim={{0.00000000\width} {0.00000000\height} {0.00000000\width} {0.08357698\height}},width=\columnwidth,clip]{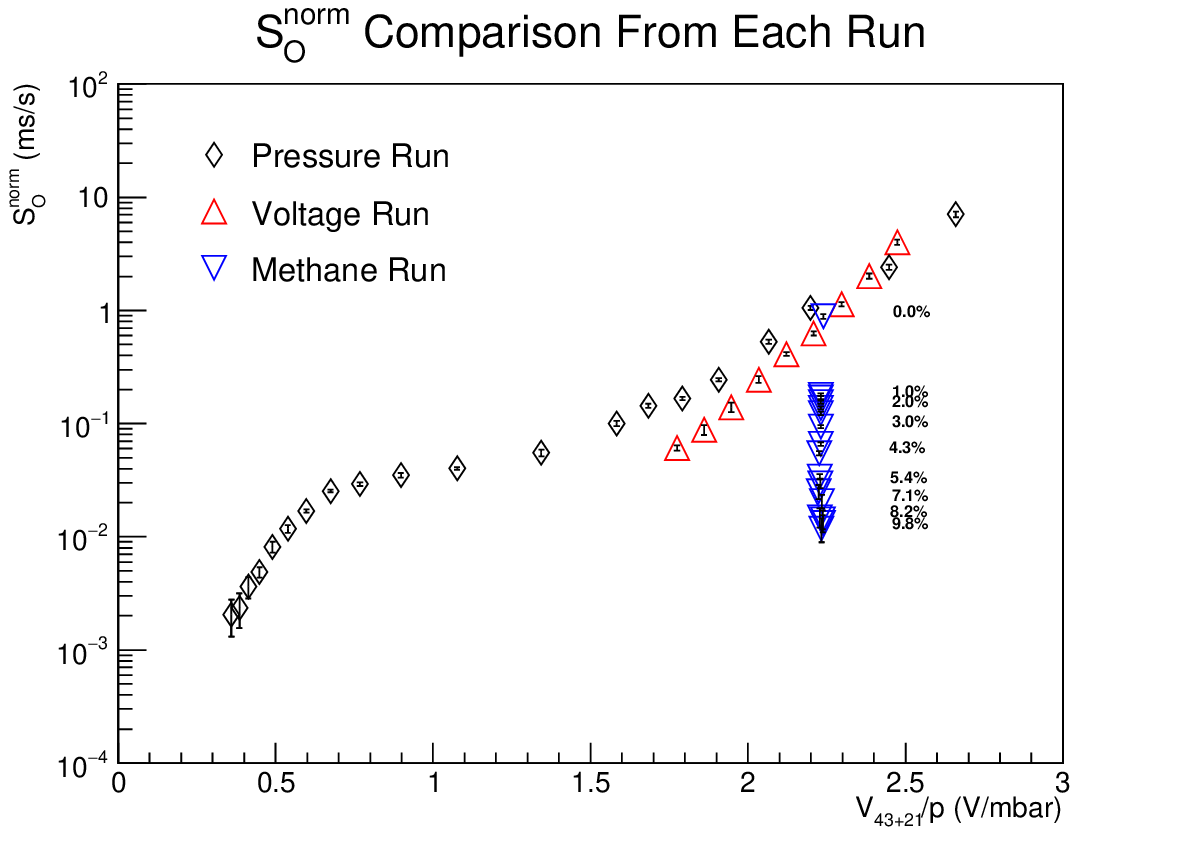}
    \caption{Comparison of $S_{\mathrm{O}}^{\mathrm{norm}}$ as a function of $V_{43+21}/p$ for the \textit{voltage}, \textit{pressure} and \textit{methane} runs shown in Figs.~\ref{fig:volt-opt-lin}-\ref{fig:met-opt-lin}. The methane fractions are indicated next to the data points.}
    \label{fig:xmas-comp-opt}
\end{figure}

The results of the three commissioning runs are compared in Fig.~\ref{fig:xmas-comp-opt}. The pressure-dependent deviation from an exponential decrease, away from what a decreasing $V_{43+21}/p$ can account for, requires further theoretical investigation and  follow-up measurements. Such studies could clarify the detailed light production mechanism in argon gas. 

\section{Summary}
\label{sec:5-conclusion}

Precision measurements of neutrino interactions in the GeV regime require  next-generation detectors. Active-target TPCs have advantages in  detection threshold and acceptance. An appropriate gas mixture could reduce or remove the inherent systematic uncertainties in neutrino detection due to nuclear effects. With high-pressure gas in a large volume, the limited target mass could be further compensated  by an intense neutrino flux. For large volumes or surface areas, readout technologies need to be optimised   for costs and performance. The WarTPC platform was established for R\&D in these directions. It is a challenge to design a single system that fulfils all these requirements but also an opportunity to explore potential applications of each technology. 

This paper reports the construction and commissioning of WarTPC. We present the design of the pressure vessel, TPC, the camera system, as well as the slow-control, DAQ, high-voltage power supply, and electronics systems. Their integration and commissioning are documented. 

The first milestone is the observation of tracks produced by  alpha particles from an $^{241}$Am source and by cosmic rays. A quantitative account of the track reconstruction will be presented in a subsequent publication. The second milestone is the measurement of the optical response at variable gas pressure and THGEM voltages in different argon--methane mixtures. A suppression of optical signal strength is observed at high pressure and with increasing methane fraction. This observation is consistent with measurements previously reported  in Refs.~\cite{Henriques2019, Neutral_bremsstrahlung_and_excimer_electroluminescence_in-noble_gases_and_its_relevance_to_two_phase_dark_matter_detectors}. These  baseline measurements demonstrate that  WarTPC  is a functioning R\&D platform towards a consolidated high-pressure TPC design for future neutrino detections.

\section*{Acknowledgements}

We thank Y.~Ramachers, G.~Bell, and K.~Mavrokoridis for helpful discussions. This work was supported by the UK Science and Technology Facilities Council (STFC) through grants \href{https://gtr.ukri.org/projects?ref=ST\%2FW005727\%2F1}{ST/W005727/1} and \href{https://gtr.ukri.org/projects?ref=ST\%2FW000571\%2F1}{ST/W000571/1}.

\printcredits

\AtNextBibliography{\small}
\sloppy
\printbibliography[heading=bibintoc]

\end{document}